\pdfoutput=1
\documentclass[prX,english,superscriptaddress]{revtex4}
\newcommand{\halfwidth}{7cm}
\newcommand{\wholewidth}{14cm}
\newcommand{\threehighwidth}{6cm}
\newcommand{\twohighwidth}{10cm}

\usepackage[utf8]{inputenc}
\usepackage{grffile}
\usepackage{graphicx}
\usepackage{subfigure}
\usepackage{amsfonts}
\usepackage{verbatim}

\usepackage{color}

\usepackage[normalem]{ulem}

\newcommand{\pin}{\ensuremath{p_\mathrm{in}}}
\newcommand{\pout}{\ensuremath{p_\mathrm{out}}}
\newcommand{\kin}{\ensuremath{k_\mathrm{in}}}
\newcommand{\kout}{\ensuremath{k_\mathrm{out}}}
\newcommand{\rhoin}{\ensuremath{\rho_\mathrm{in}}}

\begin{document}
\title{Algorithm independent bounds on community detection problems
  and associated transitions in stochastic block model graphs}
\author{Richard K. Darst}
\email{richard.darst@aalto.fi}
\affiliation{Department of Chemistry, Columbia University, 3000
  Broadway, New York, NY 10027, USA.}
\affiliation{Dept. of Biomedical Engineering and Computational Science,
P.O.Box 12200, FI-00076 Aalto, Finland}
\author{David R. Reichman}
\affiliation{Department of Chemistry, Columbia University, 3000
  Broadway, New York, NY 10027, USA.}
\author{Peter Ronhovde}
\author{Zohar Nussinov}
\email{zohar@wuphys.wustl.edu}
\affiliation{Department of Physics, Washington University in St. Louis,
Campus Box 1105, 1 Brookings Drive, St. Louis, MO 63130, USA}
\begin{abstract}
  We derive rigorous bounds for well-defined community structure in
  complex networks for a stochastic block model (SBM) benchmark.  In
  particular, we analyze the effect of inter-community ``noise''
  (inter-community edges) on any ``community detection''
  algorithm's ability to correctly group nodes assigned to a planted
  partition, a problem which has been proven to be NP complete in a standard rendition.
  Our result does not rely on the use of any one particular algorithm nor on the analysis of the limitations of inference. 
  Rather, we turn the problem on its head and work backwards to examine when, in the first place, well defined structure may exist in SBMs.
  The method that we introduce here could potentially be applied to other computational problems.  The objective of community detection algorithms is to
  partition a given network into optimally disjoint subgraphs (or {\it
    communities}). Similar to $k-$SAT and other combinatorial optimization problems,
  ``community detection'' exhibits different phases. Networks that
  lie in the ``unsolvable phase'' lack well-defined structure and thus have no
  partition that is meaningful. Solvable systems splinter into two
  disparate phases: those in the ``hard'' phase and those in the ``easy''
  phase. As befits its name, within the easy phase, a partition is
  easy to achieve by known algorithms. When a network lies in the hard
  phase, it still has an underlying structure yet finding a meaningful
  partition which can be checked in polynomial time 
  requires an exhaustive computational effort that rapidly
  increases with the size of the graph. 
  When taken together, (i) the rigorous results that
  we report here on when graphs have an underlying structure and (ii) recent results
  concerning the limits of rather general algorithms, suggest
  bounds on the hard phase. 
\end{abstract}
\maketitle

\section{Introduction}

Increasingly, data are being generated in the form of networks, where
interactions among objects are the focus of study.  Social
networks are perhaps the most prototypical example which consist of
people (nodes) and their associations (edges).  Finding structure in
complex networks is a problem of broad interest with applications in
social, biological, communications systems and many other branches.  Generally
speaking, ``community detection''
\cite{fortunato2010community,newman2004finding} attempts to identify
relevant structure in a complex network by searching for clusters of
nodes (which are termed {\it communities}) that have a higher density of
internal edges (i.e., intra-community links) than they have with other
communities (inter-community links)\cite{darst2013edge}.

A wide variety of methods for community detection have been developed over the past decade
\cite{fortunato2010community,newman2004finding}.  More recently, an
intense effort has been expended on understanding the theoretical
foundations of these methods and of community structure in
general. %\cite{need_several_examples_???}.
This was underscored by Fortunato and Barthélemy when they
demonstrated that maximizing ``modularity'' \cite{newman2004finding}, a
common measure of network partitioning, suffered from a fundamental
limitation.  Modularity is a global network parameter which measures
the quality of any particular network partition. Higher
modularity is taken to mean that more meaningful communities are found
\cite{newman2004finding, newman2006modularity}.  A fundamental
shortcoming of this method is that the local community partitions
determined by maximizing modularity depend on the global size of the
network \cite{fortunato2007resolution,lancichinetti2011limits}.

\begin{figure}[h]
  \centering
  \includegraphics{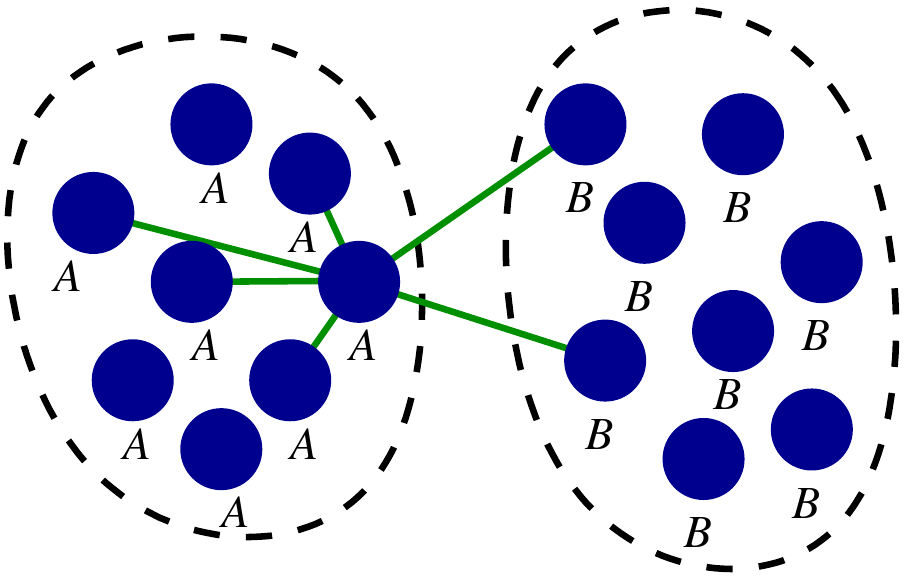}
  \caption{Planted communities.  Each node is labeled with its planted
community.  Example edges of one node of community $A$ is drawn, and
we see it has more edges to community $A$ than $B$.  Community
detection algorithms are thus expected to be able to find that this
node belongs to the $A$ community.}
\label{fig:planted}
\end{figure}

A prevalent method of judging the performance of community detection
algorithms involves ``planting'' initially well-defined community
partitions into a random network in progressively more challenging
contexts (more extraneous and/or fewer inter-community edges). 
A vivid example of such a planted state is provided by the cartoon of Fig.~\ref{fig:planted}
depicting $q=2$ communities, each with $n=8$ nodes, and the internal/external edges associated with a specific
node; in this cartoon for each node
there are far more internal intra-community edges than links between nodes in different communities. 
Such benchmarks are generally defined by a set of parameters specifying the
edge densities within the planted communities and an amount of
``noise'' representing additional spurious edges between nodes in
different communities.  The goal of community detection algorithms, as
applied to the particular case of benchmark graphs, is to rediscover
the embedded communities in given network with no prior information
about the planted partition.  Community detection methods are then
tested against increasingly challenging benchmarks until detection of
viable communities becomes impossible.  Common benchmarks of planted
partitions include stochastic block models, a benchmark by
Lancichinetti, Fortunato, and Raddicchi (LFR) which focuses on power
law distributions of nodes and degrees
\cite{lancichinetti2008benchmark}, and a variety of ``real-life
networks''.  % \revise{cite some?}
Recent theoretical work studied the limits of various community
detection methods with increasing levels of
noise.  % \revise{cite some?}

In \cite{hu2011phase}, it was demonstrated that community detection as
ascertained by Potts models in rather general random power-law type
graphs may exhibit sharp spin-glass type transitions as a function of
increasing noise in the limit of large system size. These transitions
were evinced via thermodynamic functions, dynamical quantities, and
information
theoretic overlaps.  Various groups have obtained highly noteworthy
explicit equations for the
transition lines for the community detection problem as applied a
particular class of graph, the uniform (or vanishing power law)
``stochastic                   
block model'' (SBM) on which we further focus on in this paper.  As we
will elaborate on, SBMs are random graphs defined by a constant number
$n$ of nodes per community and fixed intra- and inter-community edge
densities. SBMs exhibit a transition whose
presence was made very evident in numerical studies of a Potts type
approach to community detection as the system size was progressively
increased, e.g., \cite{ronhovde2009multiresolution}.  We note the
analysis of Decelle, \emph{et. al.} (DKMZ)
\cite{decelle2011inference} on the inference of possible assignments
of nodes to their proper communities in sparse (i.e.,
graphs with a small
number of links per node) SBM networks which has led to an explicit
equation
for the detectability threshold.  Later stringent results by Mossel,
Neeman, and Sly \cite{mossel2012stochastic} on the general limits of
Bayesian inference for $q=2$ component sparse graphs concurred 
(as a bound) with the formula derived by DKMZ via their cavity-type approximations. 
In an illuminating work, Nadakuditi and Newman (NN) \cite{nadakuditi2012graph}
reported on an inherent accuracy limit for modularity-based community
detection methods as appplied to SBMs.  NN found an
expression identical to the earlier result of DKMZ for the threshold of noise beyond
which spectral methods cannot resolve the communities.  In the
analysis of NN, average degree diverges with the network size.  After
comparing to the earlier result of DKMZ, Nadakuditi and Newman
asserted that no method can perform better than modularity on SBM
graphs.
NN further provided an important additional formula for
the {\it fraction of properly detected nodes} as a function of noise.
It is this formula of NN 
 that largely inspired the current investigation. In what will follow
in this article, we will derive lower bounds for such a fraction. Earlier pioneering works
are also extremely noteworthy.  Amongst others, earlier analysis
\cite{mcsherry2001spectral} was performed for non-dilute graphs.
Based on the cavity approximation, Reichardt and Leone insightfully initially
suggested a similar threshold beyond which detection of communities is
no longer possible \cite{reichardt2008detectable}.

In the present work, we will derive universal rigorous bounds on
the feasibility of well-defined community structure independent of any specific community
detection methods.  Our results are related to fundamental, generic
properties of community structure and pertain to {\it general networks}. {\it We rigorously examine when a benchmark
network may have any meaningful structure in the first place.}
We do this by looking microscopically at each node to determine the
ability of any algorithm to properly classify that node: an assertion
that each node be more strongly connected (by more edges) to its
intended community than to any other community.  Due to stochastic
fluctuations in edge placement, there typically is a nontrivial (and as
we illustrate in this work, computable) number of nodes where this is not the case.  We will derive
rigorous upper bounds for the fraction of nodes which may be
detectable at all.  While these calculations lead to an upper bound, they
are \textit{not} necessarily equivalent to a calculation of optimal
performance of community detection algorithms.  In fact, we find
further evidence for a region where graph structure exists
but certain methods may not be able to detect it.

We now elaborate in more detail on the stochastic block
benchmark and further general reasoning behind benchmark graphs.
Although the ideas advanced in our work hold for any system, we will,
for concreteness, focus on the stochastic block model (SBM).
The SBM is defined as follows. Each node belongs to one community (or ``block'')
\cite{holland1983stochastic,heimlicher2012community}. The probability of having an edge
between two nodes depends solely on their community
membership. Specifically, the probability ($p_\mathrm{out}$) of having an edge between
any pair of nodes belonging to the same community (intra-community edges) is fixed to
a certain value $p_\mathrm{in}$ and the probability of having an
edge between any two nodes that belong to two different blocks
(inter-community edges) is given by another uniform value
$p_\mathrm{out}$.

When a benchmark graph is used, it is assumed that there is a planted
community structure which one may (and should if the community detection algorithm is good) consistently detect. Such a planted
structure can be thought of being generated by the following gedanken
experiment. Imagine that we ``plant'' a community partition solution by
dividing a group of $N$ nodes into $q$ equivalent, completely decoupled, communities
of size $n \equiv N/q$ each. Within each community the probability that a given
node will be connected by an edge to any other node in the same
community is $p_\mathrm{in}>0$.  An absence of any inter-community
links in the initial decoupled state implies that the probability of having a link between any two
nodes that belong to different communities is $p_\mathrm{out} =0$.  In
the absence of any such inter-community links, finding the planted
communities is a relatively easy task. Next, imagine that more and
more edges are added between different communities 
(i.e., that the  probability for outside links (or ``noise'' ) $p_\mathrm{out}$ is progressively increased) while the number of
intra-community edges is left unchanged. It is intuitively clear that
for ``small'' $p_\mathrm{out}$, finding the planted community will
be easy while for sufficiently large $p_\mathrm{out}$, the structure
of the planted partition will be no longer be well-defined.
Previous work shows that the network structure becomes hard to detect at
the point $  p_\mathrm{in}-p_\mathrm{out} = \sqrt{\frac{1}{n}
  \left(p_\mathrm{in} + (q-1)p_\mathrm{out}\right)  }$, which is
distinctly different from the point of loss of structure at
$\pin=\pout$\cite{decelle2011inference, nadakuditi2012graph}.  This
implies that there are points where there is some nominal structure
present in the graph ($\pout<\pin$) yet that structure is undetectable
by any method.  Instead of focusing on detectability directly, we
study the amount of structure which is actually present in the
graph.

When the planted communities are, as is the case for sufficiently small
$p_\mathrm{out}$, reflected in the edge structure, we will term these graphs
``well-defined.''  An ``ill-defined'' graph is, for a sufficiently large
$p_\mathrm{out}$, one in which the assigned edge structure does not
reflect the planted communities.  An extreme example of an ill-defined
graph is afforded by a network in which nodes are assigned (planted) into
communities, but then all edges are randomly defined with no
preference for intracommunity or inter-community links.  No community
detection method would be able to detect the purported ``communities''
in this case as the input (edge structure) is assigned independent
of the intended communities.
In the above description, well-defined structure is described as a
graph property, but it can also be applied to individual nodes.
We reiterate that our focus is not simply that of establishing a
community detection limit.  Rather, we focus on a fundamental limit
concerning the creation of such benchmark graphs.  Several authors
have hinted at this effect in the past, however, it has never been
rigorously analyzed and considered as a fundamental limitation of
community detection
\cite{lancichinetti2008benchmark,lancichinetti2009benchmarks}.  While
we do not directly explain the origin of the detectability threshold
in this work, we provide tools to analyze the divergence of structure
and detectability.

\section{Outline}

The remainder of this work is organized as follows: We start our
discussion with the definition of the SBM problem and our criterion
for a well defined community partition. In section \ref{newton_binom},
we set up our formalism and express the problem of a planted state in
terms of a binomial distribution.  In section \ref{BM-community}, we
write the corresponding exact expressions for the probability that a
planted partition satisfies our criterion for a well defined community
partition.  We then introduce, in section \ref{sec:independence_approx}, a
simple approximation for computing these probabilities. In section
\ref{fraction}, we write general exact expressions for the fraction of
nodes in the planted partition that satisfy our criterion for
community detection. We then turn to illustrate via a trivial
application of Jensen's inequality that the approximation of section
\ref{jensen} provides rigorous lower bounds
on the probability that our criteria for a well defined partition are
satisfied.
Armed with all of these formal results for the SBM
problem, we then briefly turn to consider their implications.  We
first consider, in section \ref{sec:thresholds}, threshold values of the
noise $p_\mathrm{out}$ beyond which well defined community partitions are no
longer possible. This will allow for a relation between our results
and the phase boundaries of the detectable region of the phase
diagram. In section \ref{sec:consequences}, we discuss the abstract
meaning of our ``well-defined fraction'' and make a comparison,
between our analysis and that of systems in which ill-defined nodes are shifted
to their correct communities. We then discuss, in section \ref{apply}, the breadth of graph
types and community detection methods to which our work is applicable.
 In section \ref{sec:phase_transitions}, we examine the behavior of our well
defined transitions as our systems increase in size; we will see that our
transitions between well- and ill-defined
benchmarks become sharp.  In section \ref{sec:NN_compare}, we compare
our limits of well-definedness to other established limits of
community detection to see that they agree in a certain limit,
with the a region of well-defined but undetectable communities
shrinking as our system grows larger. we point out,
in section \ref{sec:BM-accurate_methods}, how high performance algorithms lead to results that coincide, in certain limits, with our computed ``well-defined'' fraction.
In section \ref{general_section} we discuss possible extensions of our approach to 
other problems and further note possible bounds on the region
where a computationally solvable ``hard phase'' appears; in this phase, 
purported solutions of the community detection problem
may be easily checkable in polynomial time (as in any NP problem\cite{fortnow2009status})
yet finding these solutions (even though they exist) might not be efficiently achieved by general algorithms.
We conclude, in section \ref{sec:BM-conclusions}, by summarizing our main results with an eye towards their practical significance.

\section{Binomial distributions of edge densities}
\label{newton_binom}

As we stated in the Introduction, when formally defined, SBMs constitute benchmark
graphs in which planted communities are specified for each node, and
edges are assigned between every pair of nodes with a probability
$p_{AB}$ which depends only on the communities $A$ and $B$ of the two
nodes, respectively.  In order for communities to be defined, one
would expect that the density of links between communities inside a
community $A$ will exceed that between communities $A$ and
$B$\cite{darst2013edge},
\begin{equation}
  p_\mathrm{in} = p_{AA} > p_{A, B\neq A} = p_\mathrm{out}.
  \label{eq:palphabeta}
\end{equation}
However, we reiterate that in this work, we will show that due to the
fluctuations in number of edges connecting a given node to other
communities (\emph{i.e.}, not its \emph{own}), it is possible, in fact
likely, to have nodes which are not well-defined in their community
even with $p_{AA}$ above the threshold of Eq.~(\ref{eq:palphabeta}).
Unlike some other particular approaches, 
\cite{nadakuditi2012graph,decelle2011inference,mossel2012stochastic,mcsherry2001spectral,reichardt2008detectable,hu2011phase}
this is not a limit of a particular community detection method, or limitation on possible inference of structure.
Amongst earlier works on SBMs that invoke inference methods, we explicitly re-iterate and note anew the
cavity-type approximations of 
\cite{decelle2011inference} and a later non-trivial rigorous result \cite{mossel2012stochastic}
partially reaffirming the cavity approximations for when structure in sparse SBMs may be not be inferable for the 
particular case of $q=2$ communities. Instead of asking whether inference or other methods may succeed, our results relate to the {\it fundamental structure} of the graph itself, as suggested in \cite{lancichinetti2008benchmark,lancichinetti2009benchmarks}.  As
such, our results lead to universal intrinsic
bounds for any community detection method or inference considerations.

We now turn to the SBM network with $q$ communities of
size $n_A$ nodes per community $A$ and a total number of nodes $N=\sum
n_A$, with particular values of $p_\mathrm{in}$ and $p_\mathrm{out}$.
In a random realization of an SBM graph 
with these parameters, the internal degree
$k_\mathrm{in}$ of a node to its community $A$ (the ``internal degree'') follows the binomial
distribution ($\mathcal{B}$) with $n_{A}-1$ attempts to make an edge, each
with a probability $p_\mathrm{in}$,
\begin{equation}
  P[k_\mathrm{in}=k] = \mathcal{B}(k ; n_{A}-1, p_\mathrm{in}).
\end{equation}
Associated with any particular \emph{single} external community $B$, there is a binomial
probability distribution of the external degree
\begin{equation}
  P[k_{\mathrm{out}, B}=k] = \mathcal{B}(k ; n, p_\mathrm{out}).
\end{equation}
The form of the binomial distribution is given at the end of this section.
Note that $k_{\mathrm{out}, B}$ is the external degree to a specific community $B$, as
opposed to the external degree summed over all $(q-1)$ external
communities.  Also, note that there are $n-1$ possible internal edges
since we consider nodes to not link to themselves, but $n$
possible external edges to each external community.  

If all communities are of the same size $n$ (i.e., if all communities are comprised of $n$ nodes),
we will say that  
a node $a$ ``is
well-defined'' in its planted community $A$ for this node if there are more internal than
external connections \emph{to any one external community}. That is,
\begin{equation}
  \label{eq:kDefinedCriteria}
  k_{\mathrm{out},B} < k_{\mathrm{in}}, % \forall B \neq A
\end{equation}
for {\it all} communities $B\neq A$.  One key concept here which bears
restatement is that any node can have more external (summed over the
other $q-1$ communities) than internal edges and yet still be properly
defined in its ground-state community if the many external links are
spread out over enough different external communities.  We note that
this is more relaxed than the Radicchi definition of ``weak''
community in \cite{radicchi2004defining}. As numerical results
illustrate, we can detect communities well past the weak definition
\cite{ronhovde2010local}.
Zhang and Zhao also point out another intuitive case
\cite{zhang2012community,zhang2009modularity}
where the strong and weak definitions of community structure proposed
by Radicchi are both violated.

In order to generalize our considerations to unequal sized
communities, we employ the edge density $\rho = k / n$ instead of the
raw number of links\cite{darst2013edge}. For a node to be well-defined
in a community,
the edge density to that community must be greater than the edge
density to any other community.  This assumption is discussed later.
Then, instead of Eq.~(\ref{eq:kDefinedCriteria}) {\it our criteria for
  a well defined community partition are captured in terms of edge
  densities}. That is, for each node in a
community $A \neq B$,
\begin{equation}
\label{important_density}
  \rho_{\mathrm{out},B} < \rho_\mathrm{in}.
\end{equation}
The densities in Eq. (\ref{important_density})
are defined as $\rho_{\mathrm{out},B} \equiv
\frac{k_{\mathrm{out},B}}{n_B}$ and $\rho_\mathrm{in} \equiv
\frac{k_\mathrm{in}}{n_A-1}$.  Clearly, as the number of conditions of
the form of Eq. (\ref{important_density}) that need to be checked
scale linearly in the system size (i.e., linearly in the number of
individual nodes $N$ and linearly in the number of external
communities $(q-1)$). {\it Checking} a purported
partition is a polynomial in time problem (rendering the problem as
formulated by Eq.~(\ref{important_density}) to be of the NP
type). However, as in many NP problems, finding correct partition(s) in general networks that
satisfy these constraints need not be an easy polynomial problem.
(In fact, maximizing the best well known measure for community detection,
that of modularity, was shown to be NP complete
\cite{brandes2008modularity,brandes2007finding}.)
As we discussed above, the degrees $k_\mathrm{in}$ and $\{k_{\mathrm{out},B}\}_{B \neq
  A}$ for each individual node follow a binomial distribution.
Therefore, the distributions of internal and external edge densities
are described as
\begin{eqnarray}
  P\left[\rho_{\mathrm{in}}=\frac{k}{n_A-1}\right]
    &=& \mathcal{B}(k ; n_A-1, p_{\mathrm{in}}),
    \label{eq:PinBinomial}\\
  P\left[\rho_{\mathrm{out}}=\frac{k}{n_B}\right]
    &=& \mathcal{B}(k ; n_B, p_{\mathrm{out}}),
    \label{eq:PoutBinomial}
\end{eqnarray}
for $k=0, \ldots, n_A-1$ or $k=0, \ldots, n_B$ respectively.
In the above, we employed the following shorthand for (a normalized) binomial distribution,
\begin{eqnarray}
{\mathcal{B}(m;n,p)} = { {n} \choose {m}} p^{m} (1-p)^{n-m}.
\end{eqnarray}
Away from the dilute limit, we may use a normalized Gaussian to approximate the
binomial distribution $\mathcal{B}(m;n, p) %\approx
    \approx \mathcal{N}\left(m;\left<
    x\right>, \sigma^2\right) $ where the binomial mean
is $\langle x \rangle = np$ and the variance is $\sigma^2 = np(1-p)$.
The normalized Gaussian is, explicitly, given by
\begin{eqnarray}
\mathcal{N}(x ; \left< x \right>, \sigma^{2}) = \frac{1}{\sqrt{2 \pi \sigma^{2}}} \exp[- [x - \left< x \right>]^{2}/(2 \sigma^{2})],
\end{eqnarray}
where, in the argument, $ \left< x \right>$ specifies the mean of the Gaussian and $\sigma^{2}$ its variance.
Employing the normal distribution, Eqs. (\ref{eq:PinBinomial}, \ref{eq:PoutBinomial}) read
\begin{eqnarray}
  P\left[\rho_{\mathrm{in}}=\frac{k}{n_A-1}\right] &=&
       \mathcal{N}\left(k;(n_A-1) p_{\mathrm{in}}, (n_A-1)p_{\mathrm{in}}
    (1-p_{\mathrm{in}})\right),
    \label{inp}\\
  P\left[\rho_{\mathrm{out}}=\frac{k}{n_B}\right] &=&
    \mathcal{N}\left(k;n_B p_{\mathrm{out}}, n_B p_{\mathrm{out}}
    (1-p_{\mathrm{out}})\right).
    \label{outp}
\end{eqnarray}
We can invoke the identity
$\mathcal{N}(mn;np, np(1-p)) = \mathcal{N}(m;p, {p(1-p)\over n})$ to
simplify the evaluation of these probabilities.
In sparse graphs for which $\lim_{n \to \infty} np = \lambda$
with $\lambda$ denoting a constant of order unity, %with
we can approximate the
binomial distribution by a Poisson distribution instead of a normal Gaussian distribution. 
We use the equivalence $\lim_{n\to\infty}\mathcal{B}(m;n, p) \approx
\mathrm{Pois}(m;np) =  \frac{\lambda^{m}}{m!} \exp[- \lambda]$. That is,
\begin{eqnarray}
  P[\rho_{\mathrm{in}}=x] &=&
    \mathrm{Pois}(x ; (n-1)p_\mathrm{in}), \\
  P[\rho_{\mathrm{out}}=x] &=&
    \mathrm{Pois}(x ; np_\mathrm{out}).
\end{eqnarray}
For very dense graphs ($p \approx 1$), we could conceivably use the
Poisson distribution to
model missing, rather than existing, edges.

\section{Probabilities for well defined communities}
\label{BM-community}

In view of Eqs. (\ref{eq:kDefinedCriteria}, \ref{important_density}),
our goal is to determine the probability that there are \emph{not}
more external links than internal, i.e., to compute
\begin{equation}
  P\left[\max_B({\rho_{\mathrm{out},B}}) < \rho_\mathrm{in} \right].
\end{equation}
To do this, we observe that the probability for just one external community
$B$ to have less links to a given node $a$ in $A$ than all of the links
between that given node $a$ and other nodes in the same community $A$ is,
from Eqs. (\ref{inp}, \ref{outp}), given by
\begin{equation}
  P\left[\rho_{\mathrm{out},B} < \rho_\mathrm{in}\right] =
  \int dx ~
  P\left[0 <
     \mathcal{N}\left(x;p_\mathrm{in},
         {p_\mathrm{in}(1-p_\mathrm{in}) \over n_A-1 }\right)
     - \mathcal{N}\left(x;p_\mathrm{out},
         {p_\mathrm{out}(1-p_\mathrm{out}) \over n_B}\right)
   \right]. \label{eq:pdifftransition}
\end{equation}
We can approximate the normalized probability function for the
difference between the two normal distributions in the above equation
by a normal distribution whose expectation value is given by the
difference between the respective means of the two normal
distributions and whose variance is given by the sum of the variances,
\begin{eqnarray}
  P_{w, AB} & \equiv &
 ~  P\left[ \rho_{\mathrm{out},B} < \rho_\mathrm{in}\right] =
 \int_{0}
 ^{\infty} du~
  \mathcal{N}\left(u;  p_\mathrm{in}-p_\mathrm{out},
       {p_\mathrm{in}(1-p_\mathrm{in}) \over n_A-1 } +
       {p_\mathrm{out}(1-p_\mathrm{out}) \over n_B }
       \right).
     \label{eq:indepPnormalapprox}
\end{eqnarray}
     That is,
     \begin{eqnarray}
  P_{w, AB} &=&
    \frac{1}{2}\left[1-\mathrm{erf}\left(
         \frac{p_\mathrm{out} - p_\mathrm{in}}
              {\sqrt{2}\sqrt{{p_\mathrm{in}(1-p_\mathrm{in}) \over n_A-1 }
                  + { p_\mathrm{out}(1-p_\mathrm{out}) \over n_B.}}}
        \right) \right]. \label{pwab}
\end{eqnarray}
With the aid of these probabilities, we will turn to compute the fraction of correctly identified nodes. 

\section{An independence approximation for community edge density comparisons}
\label{sec:independence_approx}

In what follows, we introduce an intuitive approximation which as we will later demonstrate
gives rise to rigorous bounds on the exact problem. In Eqs. (\ref{eq:indepPnormalapprox},\ref{pwab}), 
$P_{w,AB}$ is the probability that any given node $a$ is well defined
(``$w$'') in its planted community $A$ with respect to community $B$.
Stated alternatively, $P_{w,AB}$ is the probability of any given node in community
$A$ has more edges to nodes in a given community $B$ than to those in its own community, rendering its membership in
$A$ questionable.
In each of the $(q-1)$ external communities
$B \neq A$, there is a probability $P_{w,AB}$ of having more edges connecting nodes in $B$ to
node $a$ than edges connecting to $a$ from other nodes in its own community $A$.  \emph{If all of these
  probabilities were independent}, then the probability that any such
individual node $a$ in community $A$ is well defined with respect to
all other external communities would be
\begin{equation}
  P_{w,A}^\mathrm{ind} = \prod_{B \neq A} P_{w, AB}.  \label{eq:indepPfirst}
\end{equation}
The superscript (``$\mathrm{ind}$'') signifies that this result holds
within the approximation of {\it independence} between the various
probabilities.  If denote by $X_{w,A}$ the fraction of nodes in community $A$ which are well
defined with respect to all external communities then (generally
independent of any approximation),
the fraction of nodes properly defined in the entire graph is
\begin{equation}
  X_w = \sum \frac{n_A}{N} X_{w,A}, \label{eq:indepPgeneral}. 
\end{equation}
Within the independence approximation, 
\begin{equation}
  X_{w,A}^\mathrm{ind} = P_{w,A}^\mathrm{ind}.
  \label{XP}
\end{equation}
 For equal communities, we have within the approximation of
independence
\begin{equation}
  \label{eq:indepPconstantn}
  X_w^\mathrm{ind} =
                 \left(X_{w, AB}^\mathrm{ind}\right)^{(q-1)}.
\end{equation}
$X_w$ is the probability that any one node \textit{will} be properly
defined in its ground-state community.
Eqs.~(\ref{eq:indepPfirst}, \ref{eq:indepPconstantn}) constitute the
\emph{independence approximation}. Eqs. (\ref{eq:indepPfirst}, \ref{eq:indepPconstantn}) hold for general or
equal size communities respectively. 

\begin{figure}
  \centering
  \begin{tabular}{cc}
    (a) $q=4$ $n=32$ $p_\mathrm{in}=.1$ &
    (d) $q=4$ $n=16$ $p_\mathrm{in}=.5$ \\
  \includegraphics[width=\threehighwidth]{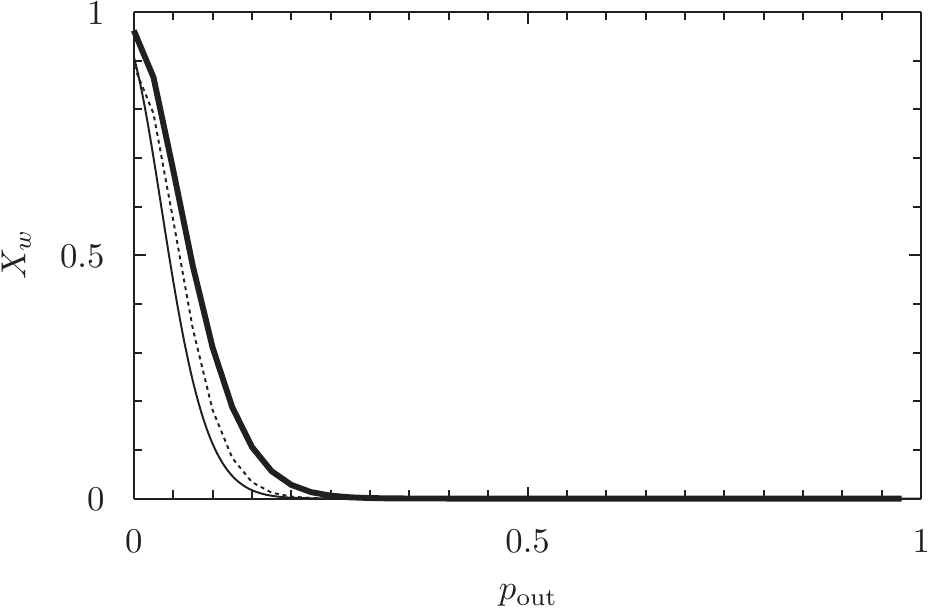} &
  \includegraphics[width=\threehighwidth]{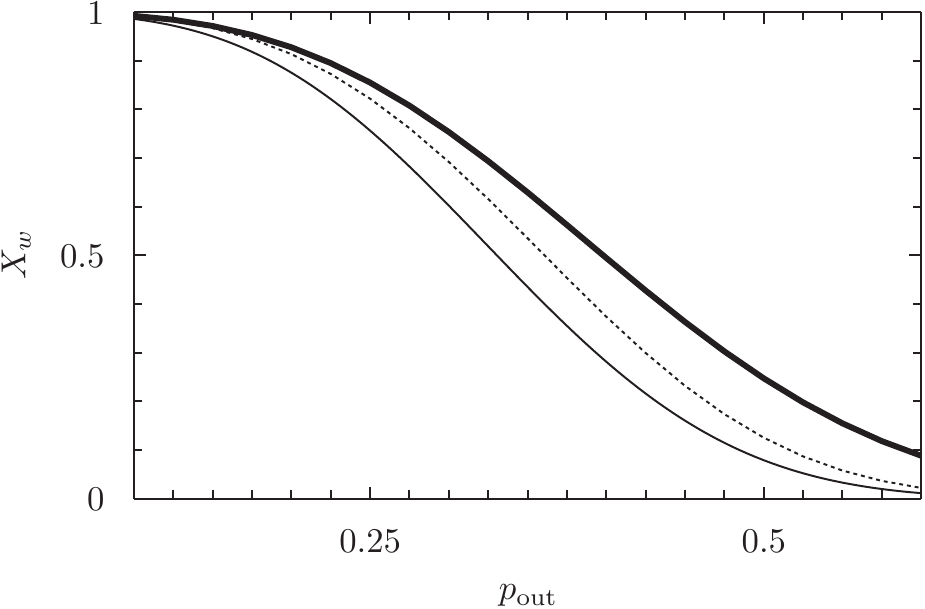} \\
    (b) $q=4$ $n=32$ $p_\mathrm{in}=.5$ &
    (e) $q=4$ $n=64$ $p_\mathrm{in}=.5$ \\
  \includegraphics[width=\threehighwidth]{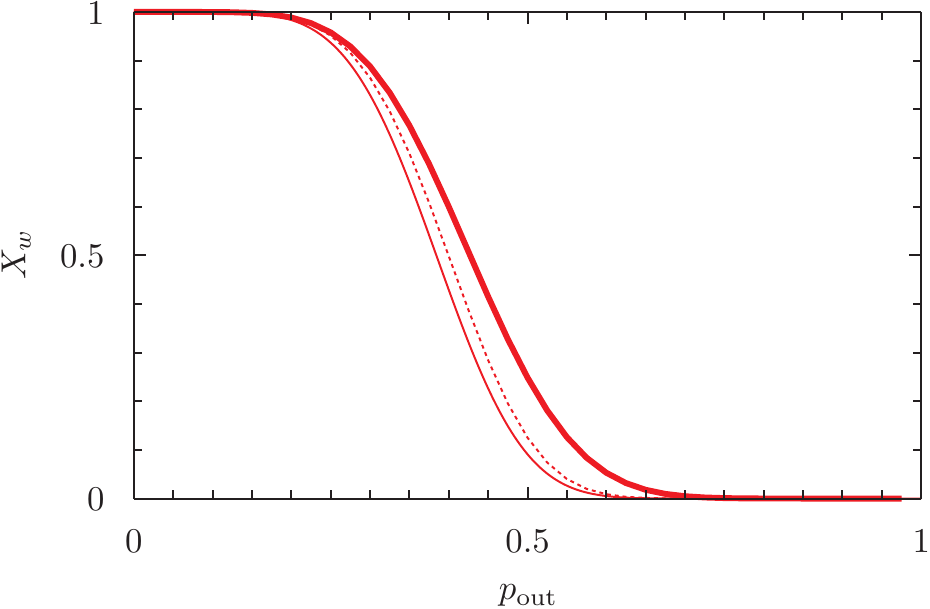} &
  \includegraphics[width=\threehighwidth]{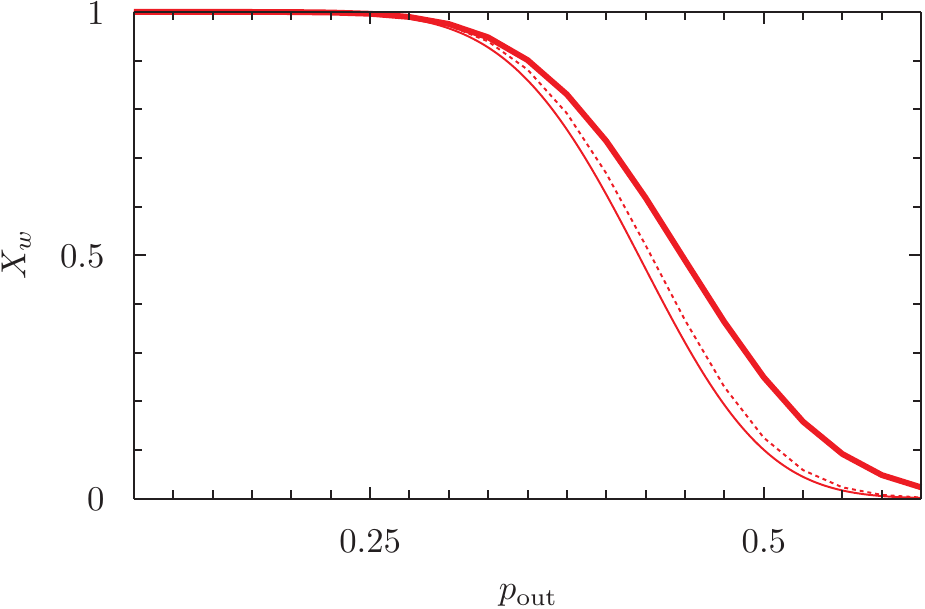} \\
    (c) $q=4$ $n=32$ $p_\mathrm{in}=.9$ &
    (f) $q=64$ $n=64$ $p_\mathrm{in}=.5$ \\
  \includegraphics[width=\threehighwidth]{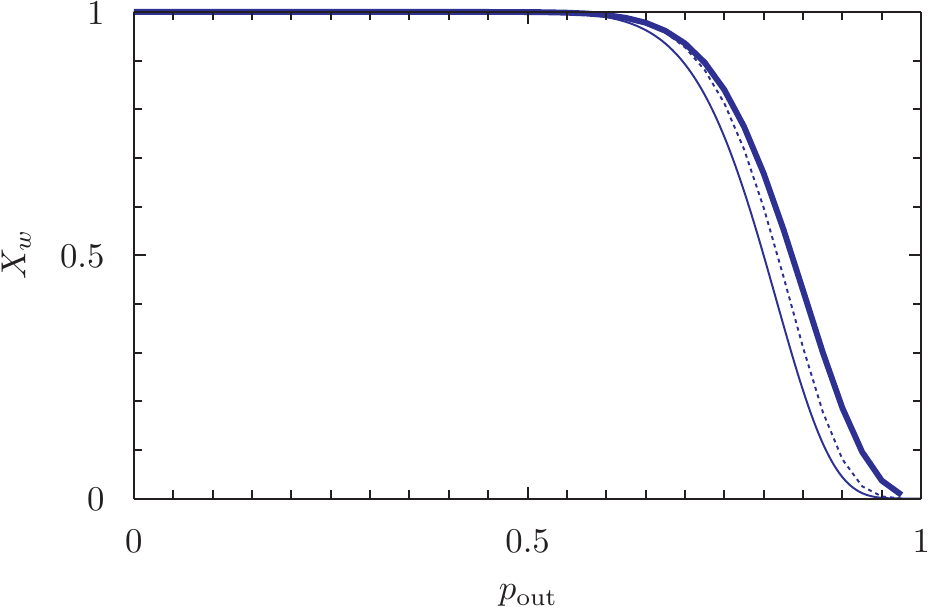} &
  \includegraphics[width=\threehighwidth]{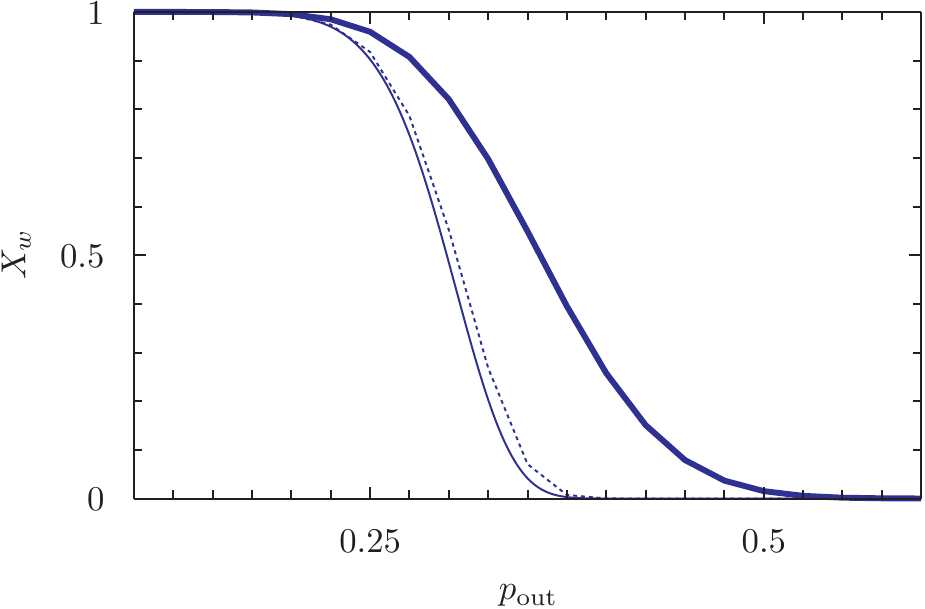} \\
  \end{tabular}
   \caption[Comparison of exact and independence approximation for
  calculation of well-definedness ]{
  Plots of the fraction $X_{w}$ of well defined nodes 
  [i.e., satisfying Eq. (\ref{important_density})] in the symmetric SBM with $q$ equal size communities 
  each having $n$ nodes as a function of 
  the ``noise'' $p_\mathrm{out}$ (the probability for any given inter-community pair of nodes to be connected by an edge)
  for different values of $p_\mathrm{in}$ (the probability for any given intra-community pair of nodes to be connected by an edge).
  The different curves provide a
  comparison between the exact calculation
    (thick line, Eqs.~(\ref{eq:PinBinomial},
    \ref{eq:PoutBinomial},\ref{eq:Pexact})), and independence approximation (thin
    solid line, Eq.~(\ref{eq:indepPconstantn})), and independence
    approximation with normal approximation (dotted line,
    Eq.~(\ref{eq:indepPconstantn}) with Eq. (\ref{pwab})).  We see
    that the independence approximation is worse for when the number
    of communities grows large or the number of nodes per community grows
    small (thick vs thin line).  The normal approximation does not
    significantly affect the accuracy of $X_w$ calculation (thin vs
    dotted line).  The region at which $X_w \approx 1$ and divergence
    from $X_w=1$ remain fairly
    well predicted under all approximations.}
  \label{fig:ApproxDiscuss}
\end{figure}

Fig. \ref{fig:ApproxDiscuss} further quantifies the validity
of the normal approximation to our problem (whose exact form is given
by the binomial distribution).  We see that adding the normal
approximation to the independence approximation (dashed vs thin solid
lines) add only a small amount of inaccuracy.

We now, as promised, turn to the
imprecision of the approximation and its origin. The fractions
$X_{w,AB}$ are not independent as assumed in
Eq.~(\ref{eq:indepPfirst}). This is so as all $X_{w,AB}$ share the
same $\rho_\mathrm{in}$.  If one particular community $B_1$ has a
$\rho_\mathrm{in}$ higher than $\rho_{\mathrm{out},B_1}$, then
$\rho_\mathrm{in}$ has a greater probability of being larger than the
average $\rho_\mathrm{in}$.  This increases the probability that a future
community $B_i$ has $\rho_\mathrm{in} > \rho_{\mathrm{out},B_i}$.
Each successive external community which has a lower $\rho_\mathrm{out} <
\rho_\mathrm{in}$ further biases the expected value of
$\rho_\mathrm{in}$ towards higher values. The overall effect is that
$X_{w,A}$ is greater than we find using the independence
approximation.  Thus, the independence approximation leads in fact
to a stringent bound,
\begin{equation}
\label{xwbound}
  X_w \ge X_w^\mathrm{ind},
\end{equation}
where the equality holds when $q=2$.  We will formally derive this
bound in section \ref{jensen}.  This general bound is vividly
illustrated by a simple example. Consider the case of $p_\mathrm{in} =
p_\mathrm{out}$ for equal size communities.  If we approximate $n
\approx n-1$, each community, including the ``in'' one, will have
equal chances of having the most connections to a given node.  Thus,
in such a case, $X_w$ should equal $1/q$.  Using the independence approximation in
Eq.~(\ref{eq:indepPconstantn}), however, we will have a $50 \%$
probability of well-definedness with respect to each external
community, thus leading to the paradoxical results that $X_w = \left(\frac{1}{2}\right)^{q-1}$ which is
flatly incorrect.  Nevertheless, this approximation gives rise to a
nontrivial bound of possible analytic utility, which we will improve
upon.  In the next section, we will compute more rigorous bounds of
the fractions of correctly identified nodes in all SBM type partitions.
We will then, as promised, illustrate that the independence approximation adheres to the bound of Eq. (\ref{xwbound}). 

\section{Fractions of well defined nodes in SBM partitions}
\label{fraction}

To produce an analytical form for the fraction of well
defined nodes $X_w$, we need to find the probability density of
the maximum of a series of random variables.  This is a hard task,
requiring nested integrals.  We can slightly simplify this problem by
considering the probability distribution 
\begin{equation}
\label{pp}
  P_{w, A}(\rho_\mathrm{in}) =
  P\left[
    \rho_\mathrm{in} > \max\{
      \rho_{\mathrm{out},B_1}, \rho_{\mathrm{out},B_2}, \ldots,
      \rho_{\mathrm{out},B_{q-1}}\}
  \right] . \label{eq:PmaxAll}
\end{equation}
 Clearly, 
\begin{equation}
\label{xwclaro}
  X_{w, A}(\rho_\mathrm{in}) =
  P\left[
     (\rho_\mathrm{in} > \rho_{\mathrm{out}, B_1}) \wedge
     (\rho_\mathrm{in} > \rho_{\mathrm{out}, B_2}) \wedge
     \ldots \wedge
     (\rho_\mathrm{in} > \rho_{\mathrm{out}, B_{q-1}})
  \right],
\end{equation}
with $\wedge$ representing the logical ``and''.  
Thus, to obtain the probability associated with the maximum,
we need to integrate the multi-variable probability distribution. 
In general, of course, there might be states other than a particular planted SBM
state for which a high fraction of well-defined nodes are found.
Eqs. (\ref{pp},\ref{xwclaro}) are however {\it independent of any particular
particular initial planted state}. As we discussed earlier and repeat here,
for any node to be well defined in a community (with $B_{1}, \cdots, B_{q-1}$ all
other external communities), it must have $
    \rho_\mathrm{in} > \max\{
      \rho_{\mathrm{out},B_1}, \rho_{\mathrm{out},B_2}, \ldots,
      \rho_{\mathrm{out},B_{q-1}}\}$, with $\rho_\mathrm{in}$
the associated link density between that node and its proper community (or communities)
that, by definition, satisfies this inequality.

We now proceed to re-examine the case of one external community
and then illustrate how to properly generalize this result for multiple 
external communities $\{B_{i}\}_{i=1}^{q-1}$. 
The probability that $\rho_{\mathrm{out}, B_i}$ is, for a {\it specific community} $B_{i}$,
less than $\rho_{\mathrm{in}}$ reads
\begin{equation}
  P\left[ \rho_{\mathrm{out}, B_i} < \rho_\mathrm{in} \right] =
  \mathrm{CDF}[\rho_{\mathrm{out},AB_i}](\rho_\mathrm{in}).
  \label{eq:pout_cumul}
\end{equation}
This reduces to Eq. (\ref{eq:pdifftransition}) in the Gaussian approximation to the binomial probability distribution function (which we return to more generally below). 
In Eq.~(\ref{eq:pout_cumul}), the
``$\mathrm{CDF}$'' syntax  denotes the ``Cumulative Distribution
Function'' of $\rho_{\mathrm{out},AB_i}$ which, when the normal approximation to the exact binomial distribution may be invoked, is given by Eqs.~(\ref{eq:indepPnormalapprox},\ref{pwab})
when these are evaluated
at $\rho_\mathrm{in}$.  Probability density functions and cumulative
distribution function nomenclature is reviewed in Appendix~\ref{sec:probdistribution}.  Generally, we may either employ the $\mathrm{CDF}$ of the
normal approximation from Eqs.~(\ref{inp},
\ref{outp}, \ref{eq:pdifftransition}) (whence, as stated above, it reduces to Eqs. (\ref{eq:indepPnormalapprox}, \ref{pwab}))
or the exact binomial from the first half of Eqs.~(\ref{eq:PinBinomial},
\ref{eq:PoutBinomial}); the normal $\mathrm{CDF}$ is, of course, a
continuous distribution, while the binomial $\mathrm{CDF}$ is a series
of discrete steps formed from a
sum of discrete quantities.  The particular form used
for the CDF reduces to an implementation detail of the necessary
numerical evaluation. However conceptually both forms (normal or binomial) can be used equivalently
with little difference in the results for most graphs (e.g., not exceptionally sparse graphs for which
the binomial distribution tends to a Poisson distribution). We now turn to the multi-community case.
From Eq. (\ref{xwclaro}),
\begin{eqnarray}
  X_{w,A}(\rho_\mathrm{in}) &=&
    P\left[ \rho_\mathrm{in} > \rho_{\mathrm{out}, B_1} \right]
    P\left[ \rho_\mathrm{in} > \rho_{\mathrm{out}, B_2} \right]
    \ldots
    P\left[ \rho_\mathrm{in} > \rho_{\mathrm{out}, B_{q-1}} \right],
  \\
  X_{w,A}(\rho_\mathrm{in}) &=&
    \prod_{i=1}^{q-1}
       \mathrm{CDF}\left[ \rho_{\mathrm{out}, B_i} \right](\rho_\mathrm{in}),
\end{eqnarray}
where the product extends over all $q-1$ external communities.  The
fraction $X_{w,A}$ of nodes which are well-defined is given by $X_{w,A} = P_{w,A}$.
Eq.~(\ref{eq:PmaxAll}) and the results above that followed apply for a given
$\rho_\mathrm{in}$.  To find out the probabilities and fraction of correctly identified nodes,
we must integrate over all possible values of
$\rho_\mathrm{in}$.
This leads to
 \begin{equation}
  X_{w,A} = P_{w,A} =
  \int_0^1
  d\rho_\mathrm{in}
  \mathrm{PDF}\left[ \rho_{\mathrm{in},A} \right](\rho_\mathrm{in})
  \prod_{i=1}^{q-1}
    \mathrm{CDF}\left[ \rho_{\mathrm{out}, B_i} \right](\rho_\mathrm{in}).
  \label{eq:Pexact}
\end{equation}

We may employ the general relation of Eq.~(\ref{eq:indepPgeneral}) to
determine the fraction of correctly identified nodes.  For
equal-size communities, we have
\begin{equation}
\label{xweqe}
  X_w =
  \int_0^1
  d\rho_\mathrm{in}
  \mathrm{PDF}\left[ \rho_{\mathrm{in},A} \right](\rho_\mathrm{in})
    \left(
      \mathrm{CDF}\left[ \rho_{\mathrm{out}, B_i}
      \right](\rho_\mathrm{in})
    \right)^{q-1}.
\end{equation}
We end up with an integral which can only be easily
evaluated for $q=2$.  For $q=2$ (and only for $q=2$), we can proceed as in
the case of Eq.~\ref{eq:indepPnormalapprox}.  Note that for $q=2$, the
independence approximation is not wrong, as there are no multiple
objects to assume independence of (e.g.,trivially, 
$\left(\frac{1}{2}\right)^{(q-1)} = \frac{1}{q}$ when $q=2$).

\begin{figure}
  \centering
  \begin{tabular}{cc}
    (a) $q=4$ $n=32$ $p_\mathrm{in}=.1$ &
    (d) $q=4$ $n=16$ $p_\mathrm{in}=.5$ \\
  \includegraphics[width=\threehighwidth]{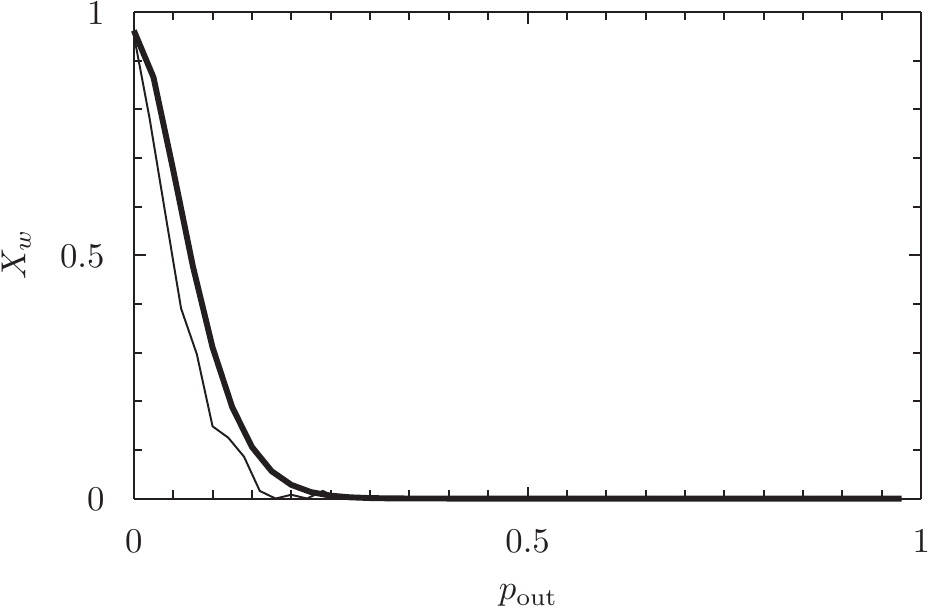} &
  \includegraphics[width=\threehighwidth]{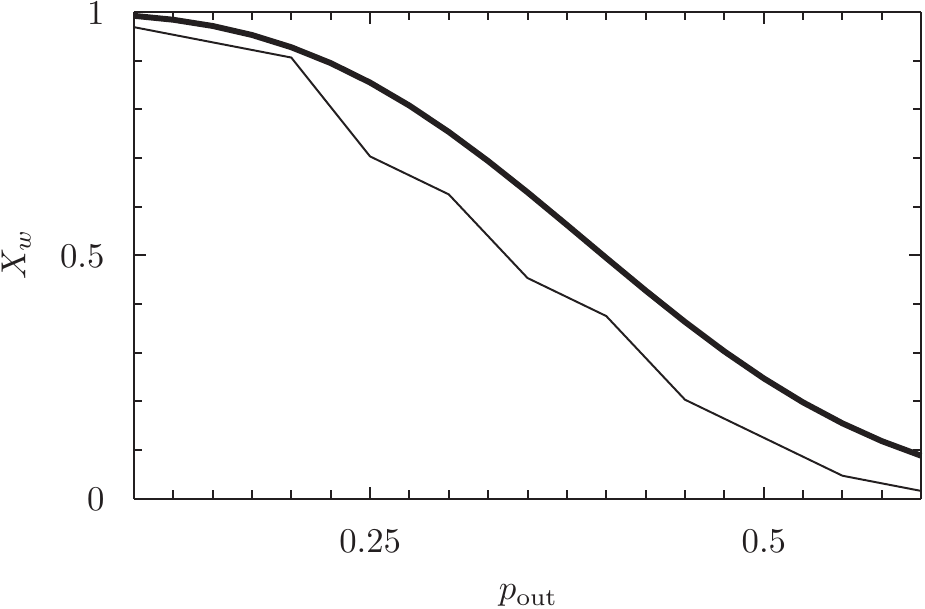} \\
    (b) $q=4$ $n=32$ $p_\mathrm{in}=.5$ &
    (e) $q=4$ $n=64$ $p_\mathrm{in}=.5$ \\
  \includegraphics[width=\threehighwidth]{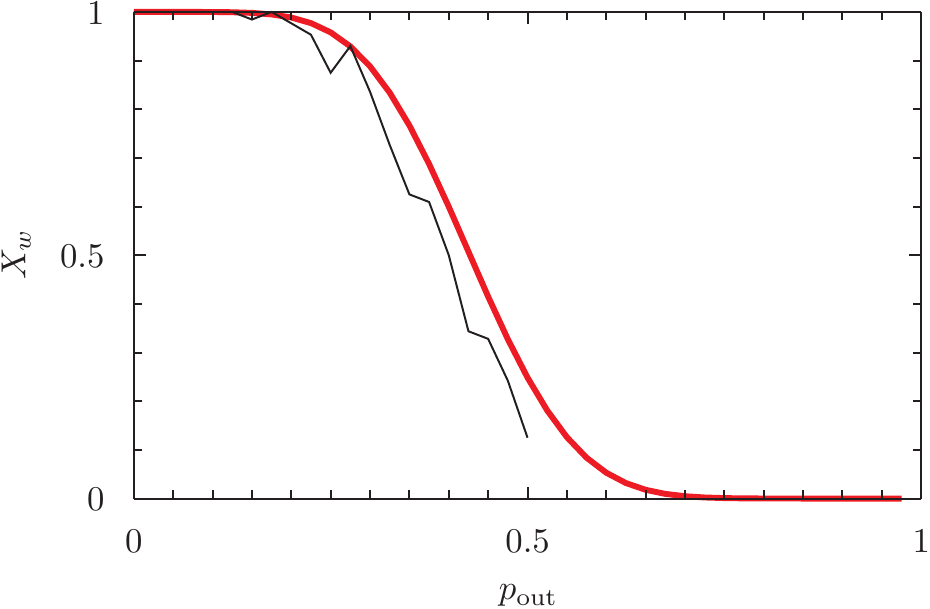} &
  \includegraphics[width=\threehighwidth]{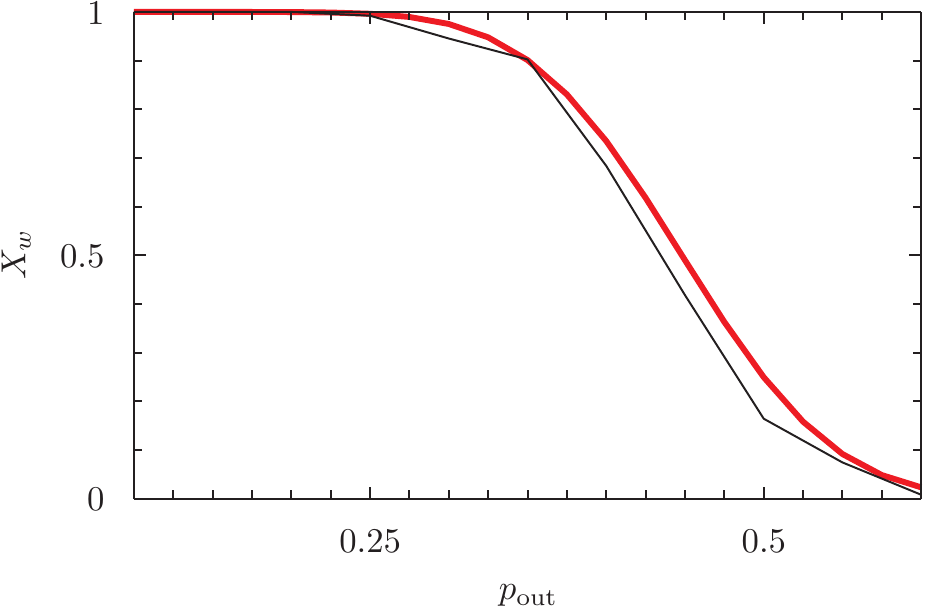} \\
    (c) $q=4$ $n=32$ $p_\mathrm{in}=.9$ &
    (f) $q=64$ $n=64$ $p_\mathrm{in}=.5$ \\
  \includegraphics[width=\threehighwidth]{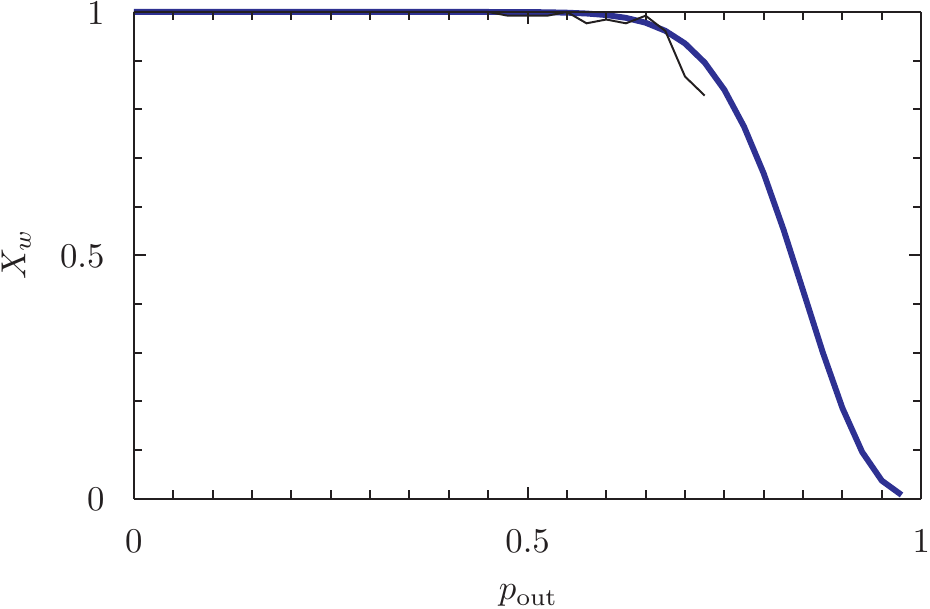} &
  \includegraphics[width=\threehighwidth]{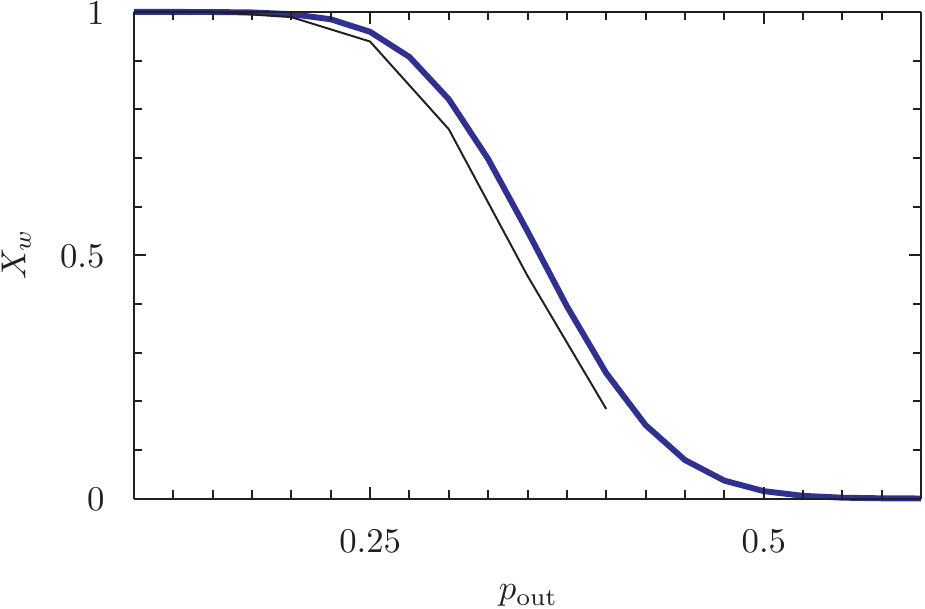} \\
  \end{tabular}
  \caption[Comparison of theoretical and experimental well-definedness
  calculation]{Comparison of exact calculation of well-defined node
    fraction $X_w$ (thick line, Eq.~(\ref{eq:Pexact},
    \ref{eq:PinBinomial}, \ref{eq:PoutBinomial})) with brute
    computation of well-definedness using
    Eq.~(\ref{important_density}) from 100 sample graph generations
    (thin line) of SBM graphs with the given parameters.  We see that
    our theoretical calculations match experimental computation well.
    However, these results do not consider \textit{cascades}: all
    nodes are assumed to stay fixed in their original communities, and
    well-definedness is calculated relative to that state.  The lack
    of cascades is the reason for a lack of a strict detectability
    threshold.}
  \label{fig:bmfitness-compareExact}
\end{figure}

That the independence approximation leading to
Eq.~(\ref{eq:indepPconstantn}) is not exact for $q>2$ is indeed evident
since, as discussed in the example which we just provided earlier, for $p_\mathrm{in} = p_\mathrm{out}$, the
approximation will yield $\left(\frac{1}{2}\right)^{(q-1)}$, while the
correct value should approach $\frac{1}{q}$.  As $q$ grows larger, the
approximation becomes less accurate and the approximation leads to
progressively less tight lower bounds on $X_w$.
Fig.~\ref{fig:bmfitness-compareExact} compares brute computation of the
well-defined fraction of nodes with our theoretical calculations.  We
see that the approximations are good as long as the number of communities
$q$ is small.

\begin{figure}
  \centering
  \includegraphics[width=\halfwidth]{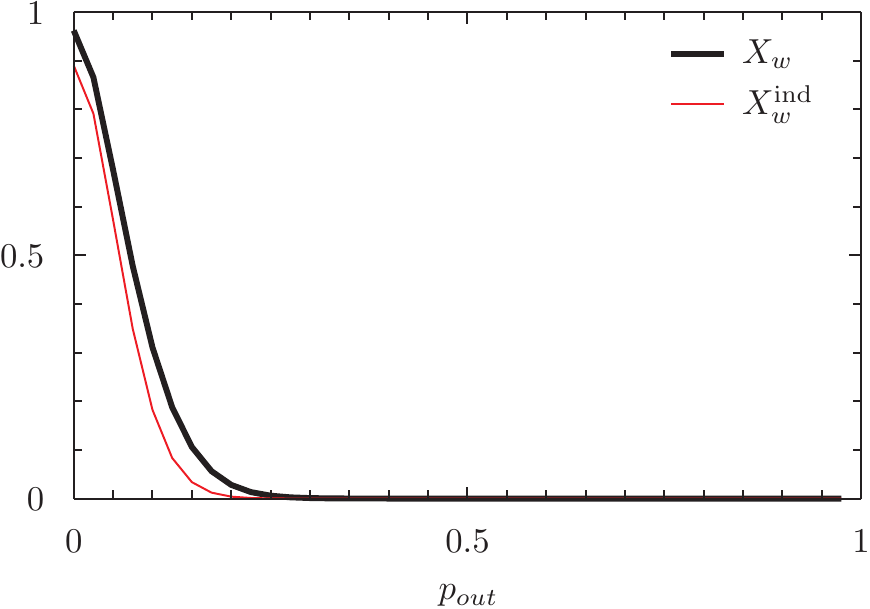} \\
  \includegraphics[width=\halfwidth]{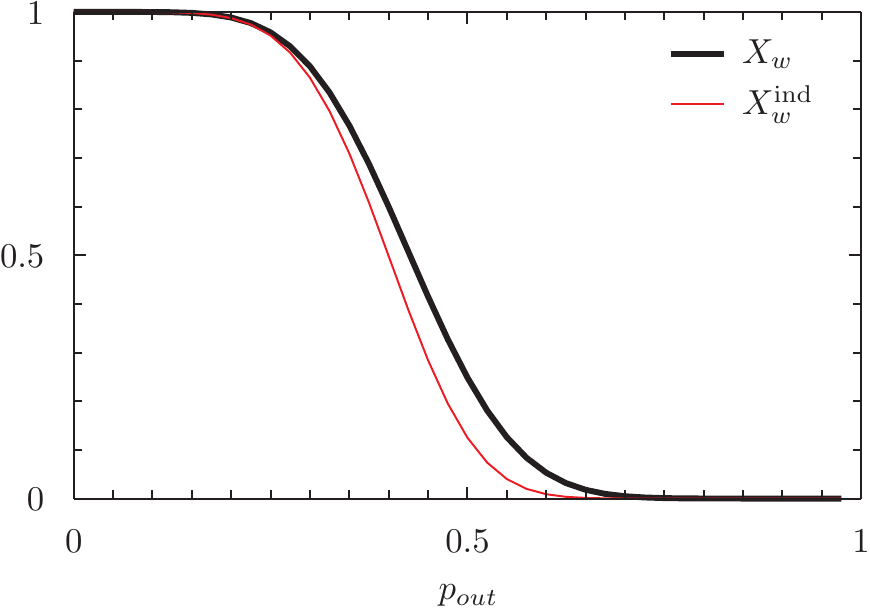} \\
  \includegraphics[width=\halfwidth]{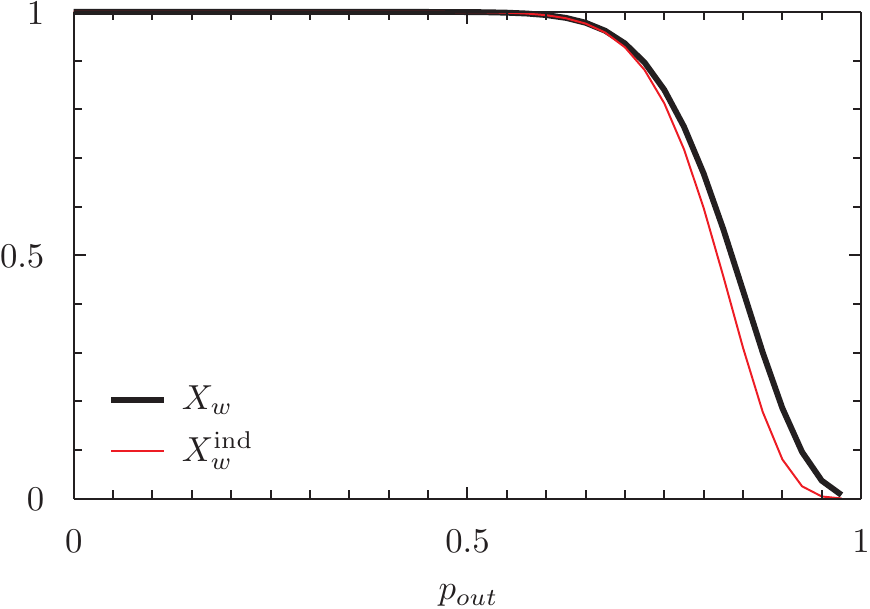}
  \caption[Comparison of well-definedness to community detection
  accuracy.]{ Comparison of different approximations for calculation
    of $X_w$ for a $q=4$ $n=32$ SBM graph at $p_\mathrm{in} = \{.1, .5,
    .9\}$ from top to bottom.  The $X_w$ calculation (thick
    line, Eq.~(\ref{eq:Pexact}, \ref{eq:PinBinomial},
    \ref{eq:PoutBinomial})) closely matches our independence
    approximation (thin solid line, Eq.~(\ref{eq:indepPconstantn})).
    Furthermore, we see that $X_w = 1$ accurately predicts the regions
    at which the
    absolute Potts model, a recently proposed accurate community
    detection method, is able to detect communities
    \cite{ronhovde2009multiresolution,ronhovde2010local}.  Once
    $X_w<1$, the community detection method rapidly loses accuracy.
  }
  \label{fig:bmfitness-normalapprox}
\end{figure}

In practice, we expect {\it cascade effects}:  Every misclassified node will
influence the number of nodes in each community, thereby affecting the
well-definedness of other nodes in those communities.  Thus, because
of these cascade effects, we expect $X_w$ to actually be smaller than
even our calculation above, with the greatest accuracy when $X_w
\approx 1$.  The fraction $X_w$ is of use in
monitoring the performance of community detection algorithms.
Fig.~\ref{fig:bmfitness-normalapprox} compares our calculation of
$X_w$ to an actual community detection algorithm, and shows that our
$X_w$ calculation accurately predicts properties of the community
detection process which will be elaborated on in future sections.

\section{The independence approximation as an explicit lower bound  on general partitions via Jensen's inequality}
\label{jensen}

Armed with the general expressions for the fraction of correctly
identified nodes $X_{w}$, we now return to explicitly demonstrate that
the independence approximation leads to explicit bounds of
Eq. (\ref{xwbound}). Within the independence approximation,
\begin{equation}
\label{xwccdfind}
  X_w^\mathrm{ind} =
  \left(
  \int_0^1
  d\rho_\mathrm{in}
  \mathrm{PDF}\left[ \rho_{\mathrm{in},A} \right](\rho_\mathrm{in})
      \mathrm{CDF}\left[ \rho_{\mathrm{out}, B_i}
      \right](\rho_\mathrm{in})
  \right)^{q-1}.
\end{equation}
Without the independence approximation, we have
the result of Eq. (\ref{xweqe}) which we rewrite here anew for 
the benefit of the reader to aid comparison,
\begin{equation}
\label{xwccdf}
  X_w =
  \int_0^1
  d\rho_\mathrm{in}
  \mathrm{PDF}\left[ \rho_{\mathrm{in},A} \right](\rho_\mathrm{in})
    \left(
      \mathrm{CDF}\left[ \rho_{\mathrm{out}, B_i}
      \right](\rho_\mathrm{in})
    \right)^{q-1}.
\end{equation}
It is readily seen by Jensen's inequality \cite{chandler1987introduction}
for convex functions $\varphi$ in its
general form as applied to general probability distribution functions
($\int dz f(z) = 1$),
\begin{eqnarray}
\label{jenseq}
\varphi\left(\int g(z) f(z) dz\right) \le \int \varphi\left(g(z)\right) f(z) dz,
\end{eqnarray}
for the particular function $\varphi(y) =
y^{q-1}$ (with $q \ge 2$), that Eqs. (\ref{xwccdfind},
\ref{xwccdf}) lead to the bound of Eq. (\ref{xwbound}).  An equality
always trivially arises when $q=2$ (as is seen from Eq. (\ref{jenseq})
for $\varphi(y) = y$). We further remark that if, {\it in addition
to all possible initially planted SBM partitions}, we also examine
non SBM type partitions of the original given graphs, 
the appearance of $X_{w}^\mathrm{ind}$
as a lower bound (Eq. (\ref{xwbound})) as we derived above 
by Jensen's inequality {\it can only be fortified}.

\section{Community Detection Thresholds}
\label{sec:thresholds}

Thus far, we have focused on the fraction of well defined
nodes $X_w$. This quantity allows us to know the maximum fraction of
nodes \emph{any} community detection algorithm can achieve.  However,
it is also useful to have a rough threshold for the case of every node
being well-defined.  Towards this end, we propose an upper threshold
probability
\begin{eqnarray}
\label{upper_threshold}
X_w^\mathrm{Th,h} = 1 - \frac{1}{N}
\end{eqnarray}
to test for a proper detection of communities.  For $X^{w}$ above $X_w^\mathrm{Th,h}$, all
nodes are well defined a majority of the time. Below
$X_w^\mathrm{Th,h}$, there is a high probability of at least one node
not being properly defined in its planted community.  At
$X_w^\mathrm{Th,h}$, exactly one node is, on average, mis-grouped.
This is the point at which we expect community detection algorithms to
no longer perfectly detect communities, and beyond this point to have
reduced accuracy.
Fig.~\ref{fig:bmfitness-2x50} indicates that this upper threshold
somewhat accurately
predicts the point at which community detection algorithms begin
losing accuracy.  For a given $p_\mathrm{in}$ value, the
$X_w^{\mathrm{Th},h}$ threshold occurs at a certain $p_\mathrm{out}$
value, which we shall indicate as $p_\mathrm{out}^{\mathrm{Th},h}$.
This threshold is important for another reason: when $p_{out} < p_\mathrm{out}^{\mathrm{Th},h}$, 
\textit{cascade effects} where ill-defined and incorrectly
detected nodes may influence the detectability of other nodes, are negligble.  Thus, 
when $p_{out} < p_\mathrm{out}^{\mathrm{Th},h}$, our calculations are expected
to be most accurate, and comparison with detectability is most valid.

At the other extreme, a lower threshold fraction of correctly grouped
nodes occurs at the point when the system is entirely decorrelated
from its expected community structure, the point at which every node
has an equal probability (of size $1/q$) to be found in any community. This
threshold fraction is defined by
\begin{eqnarray}
\label{lower_threshold}
X_w^\mathrm{Th,l} = \frac{1}{q} + \frac{1}{N}.
\end{eqnarray}
We employ a threshold of $1/q+1/N$ instead of simply $1/q$ as $X_w$ may never exactly approach the
symmetric $1/q$ value for finite size systems.
We may mark the corresponding value of $p_\mathrm{out}$ with
$p_\mathrm{out}^{l}$.  Without stochastic variance of the edge
placement, we would expect $p_\mathrm{out}^{l} = p_\mathrm{in}$.

By Eq.~(\ref{xwbound}), the results that we arrive at by the
independence approximation of Eqs. (\ref{eq:indepPfirst},
\ref{eq:indepPconstantn}) and, in particular, the corresponding values
of the threshold values $p_\mathrm{out}^{h,l}$, lead to lower
bounds. That is, if we denote by
${p_\mathrm{out}^\mathrm{ind}}^{h}$ the value of the
$p_\mathrm{out}$ for which $X_w^\mathrm{ind} = 1-1/N$ and
${p_\mathrm{out}^\mathrm{ind}}^{h}$ the value of the $p_\mathrm{out}$
for which $X_w^{\mathrm{ind}} = 1/q+1/N$ respectively, then clearly
\begin{eqnarray}
  p_\mathrm{out}^{h} &\ge& {p_\mathrm{out}^\mathrm{ind}}^{h}, \\
  p_\mathrm{out}^{l} &\ge& {p_\mathrm{out}^\mathrm{ind}}^{l}.
\end{eqnarray}
For $q>2$, the independence approximation leads to an
underestimate of the requisite noise to achieve these threshold fraction values of correctly identified nodes;  the true
critical values of $p_\mathrm{out}$ exceeds that found by the
independence approximation.

\section{The meaning of $X_w$ and the role of ill-defined nodes}
\label{sec:consequences}

In order to highlight the importance of fraction of correctly identified nodes $X_w$, we regress and note anew
how benchmark graphs are typically employed. A planted SBM is constructed and is provided
to a solver by solely providing information about all of the edges between nodes that are present in the graph. 
The solver is not, of course, told which nodes formed the different communities that were used
in the construction of the SBM. A good solver is then expected to be able to use
only the edge information to recover the planted communities.  $X_w$
is the fraction of nodes which a method classifies correctly. When $X_w = 1$, all nodes are, by fiat, properly defined, and reasonable
community detection algorithms would be expected to be able to
identify the correct community of all nodes.  However, when $X_w< 1$,
then there are some nodes which are more strongly connected to a
community other than their planted community.  In this case, \emph{no
  community detection algorithm should be expected to classify these
  nodes correctly} since the edge structure does not reflect the
planted communities. The inability to detect all correct communities might be seen as
a flaw in the algorithm, instead of a flaw in the benchmark (the basic premise of our work)
as it should be interpreted.

\begin{figure}
  \centering
  \includegraphics[width=\wholewidth]{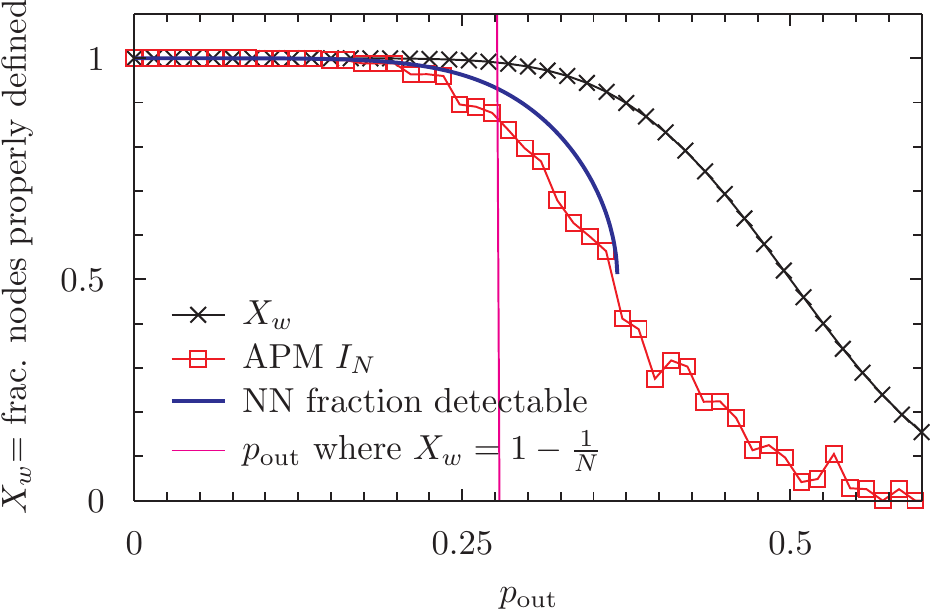}
  \caption[Comparison of experimental community detection and
  well-definedness calculations]{Comparison of $X_w$ and actual
    performance of community detection methods on a stochastic block
    model with $q=2$, $n=50$, and $p_\mathrm{in}=0.5$.  We plot $X_w$,
    a upper limit of the fraction of nodes properly classified, with
    the $X_w=1-1/N$ threshold represents the limit of a region where all
    nodes will be well-defined.  We also plot the performance of the
    absolute Potts model community detection method, and the
    ``optimal'' performance of spectral modularity-based methods is
    also plotted (Eq.~(\ref{eq:newman-detectable}),
    \cite{nadakuditi2012graph}). We note a divergence between the
    theoretical amount of structure available ($X_w$) and the
    performance of community detection methods, even below the
    $X_w=1-1/N$ threshold of all nodes being well-defined.
  }
  \label{fig:bmfitness-2x50}
\end{figure}

In Fig. \ref{fig:bmfitness-2x50}, we compare the limits of
detectability of the absolute Potts model (APM)
\cite{ronhovde2009multiresolution,ronhovde2010local} with those 
derived for general spectral methods \cite{nadakuditi2012graph}.
We observe that the upper threshold of 
Eq.~(\ref{upper_threshold}) fairly accurately predicts both (i) the point
at which the fraction of correctly identifiable nodes via spectral-based method (the NN line of 
\cite{nadakuditi2012graph}) begins decreasing, and (ii) when the APM method ceases to be able to identify communities
well.  The transitions to the undetectable phase as seen by both methods (i) and (ii) onset 
at nearly the same value of the noise $p_\mathrm{out}$.  More
interesting, however, is the fact that we see a notable divergence
between the ability to detect nodes (as calculated by NN in
\cite{nadakuditi2012graph}) and the structure present in the graph.
This is true even near our ``accurate'' range, near the low threshold
$p_\mathrm{out}^l$.

\begin{figure}
  \centering
  \includegraphics[width=\halfwidth]{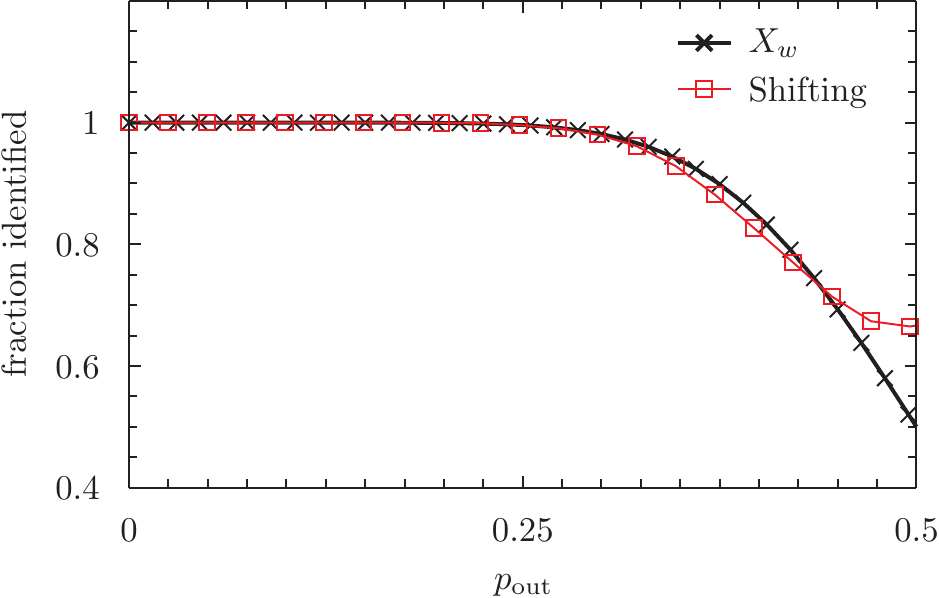}
  \caption{$X_w$ compared to a simple node shifting process in a $q=2$
    communities, community size $n=50$, $\pin=.5$ SBM averaged over
    100 graph instances.  Each ill-defined node (in the density
    formulation, Eq.~(\ref{important_density})) is taken and shifted
    to the community with which it shares the greatest edge density.
    This process is iteratively repeated until no more changes are
    possible.  We see that $X_w$ closely matches this shifting process
    for moderately high $X_w$, indicating that $X_w$ does have some
    relation to embedded graph structure, despite approximations.}
  \label{fig:bm-Xw-match-shift}
\end{figure}

We can further envision a new experiment to test the reliability of
$X_w$ in indicating the present graph structure.  We construct an SBM
graph, and each ill-defined node is shifted to its ``correct''
community, that community with which it shares the greatest number of
edges.  (In practice, we use edge density
${\rhoin}_{,C}={\kin}_{,C}/n_C$, with $n_C$ being the number of nodes
in the respective community, as a criteria for shifting).  Then, we
compare the fraction of properly classified nodes in this ``shifted''
graph to that in the planted graph.  This provides an upper bound of
available structure in the graph, if a community detection method knew
the original community structure.  However, we see
(Fig.~\ref{fig:bm-Xw-match-shift}) that there is a significant
difference between this measure and our computed fraction
$X_{w}$ in the earlier sections (as well as the
 detectability calculation of
\cite{decelle2011inference,nadakuditi2012graph}
for which $X_{w}$ is even smaller).  
This is so as these earlier results concern the structure of the
graph (and the ability to infer underlying communities).
However, in the shifting process, we begin
with the structure and slowly blur it. Consequently, the transition
to the disordered structureless phase is not as precipitous. In
Fig.~\ref{fig:bm-Xw-match-shift}, we see that $X_w$ closely
matches the fraction of nodes properly detectable by node-shifting in
the high and moderately high $X_w$ range.  While this process should
be considered unreliable for the lower threshold of $X_w \approx 1/q$, it demonstrates that
our analytic $X_w$ calculations \emph{do} capture some essence of structure
available for detection in the graph.

\section{Applicability of results}
\label{apply}

While the discussion above is focused primarily on equal-sized communities,
Eq.~(\ref{eq:indepPgeneral}) generally holds for graphs with communities
of any size.  However, once we can no longer assume equal size
communities, it becomes much more difficult to obtain analytic results. 
Numerically computing $X_w$ is still, however, a relatively
simple task.  If we go further and allow the coefficients $p_{AB}$ to
vary, the form of $X_w$ becomes more complex.  Nevertheless,
for any specific graph instance, an $X_w$ value can be calculated by
iteration over all nodes and comparing internal and external degree
and community sizes.

Our analysis invokes an edge density picture of community detection,
where a community of nodes is identified by having more internal edges
than external edges \textit{to any one other community}.  On first
appearance, this would make it seem that this analysis was specialized
to edge-density-based community detection methods.  However, given the
highly symmetric nature of constant-$n$ stochastic block models, 
edge density is the only distinguishing factor between communities.
There are other competing factors which community detection methods could use in cost
functions to judge communities, such as internal edge density, size, number
of triangles, and other higher level correlations, with each method
weighing these factors differently.  However, in SBMs with equal-sized
communities $n$, there is \emph{only} one distinguishing factor: edge
density.  Thus, all reasonable community detection methods should
converge to the same result for this special benchmark class.
  This rationalizes why various methods such as
as those invoking modularity \cite{newman2004finding}, a configuration Potts
model by Reichardt and Bornholdt \cite{reichardt2006statistical}, an
Erd\H os-R\'enyi Potts model
\cite{reichardt2004detecting,reichardt2006statistical}, its ``constant
Potts model'' extension \cite{traag2011narrow}, and an ``absolute''
Potts model (no null model definition)  \cite{ronhovde2010local} all
converge to equivalent cost functions and show equal results in the
thermodynamic limit for the constant-$n$ SBM.  Thus, Potts-type, and
possibly all \cite{decelle2011inference, reichardt2008detectable},
methods converge to the same result for this special benchmark class.
In this sense, our analysis roughly generalizes to all community
detection methods on \emph{equal size} SBM graphs.

\section{Algorithmic vs well-definedness crossovers}
\label{sec:phase_transitions}

As alluded to in the Introduction, from a practical vantage point, the
detectability limit can also be associated with a phase transition in
various cost functions (Potts-type Hamiltonian or other)
employed by real algorithms. In the current
work, we focused on a related complementary aspect-  that of well
defined structure that may be probed. Below, we
further discuss these.

\subsection{Transitions in algorithmic approaches to community detection}

Much previous work, e.g., \cite{hu2011phase,hu2011replica} has shown the existence
of several phases in community detection problems and related Hamiltonians.  
These appear to be bona fide phase transitions as system
size grows towards the thermodynamic limit.  Similar to other computational problems\cite{mezard2002analytic},
three different phases (marked (i)-(iii) below)
are discerned in community detection problems \cite{hu2011phase}. In  (i), the ``easy''
phase, community detection methods can readily detect proper communities without much effort.  
(ii) A ``hard'' phase corresponds to a region where
communities exist and are still well-defined yet due to the system complexity an exhaustive sampling is
generally required to partition the network.  (iii) In the ``undetectable'' or ``unsolvable'' phase, no clear community detection is possible
regardless of computational effort as the system lacks clear structure. 
In the spin-glass type ``absolute Potts model''  approach \cite{hu2011phase,ronhovde2010local,ronhovde2009multiresolution,hu2011replica, hu2012stability, ronhovde2012global},
the transitions between these phases are marked by both thermodynamic (and information theoretic/complexity) measures as
well as sharp dynamical spin-glass type signatures. In random graphs, clear spin glass type behavior appears in the hard phase. In the limit of progressively larger number of nodes per community $n$ in power-law graphs, 
the size of the parameter space region which supported the ``hard'' phase steadily decreased \cite{hu2011phase} suggesting that this phase
might disappear in the large $n$ limit. The existence of these phases and their physical content is made visible in some applications such as image
segmentation \cite{hu2011replica} and a graph theory based analysis of the structure of glass formers \cite{ronhovde2011detecting, ronhovde2012detection}. In a companion paper, we illustrate how our edge density based criteria for community
detection (in particular that of Eq. (\ref{important_density})) naturally coincide with a general edge density based framework for community detection that includes the absolute Potts model.
A mechanical system that can be derived on general graphs 
which forms a continuous dual to the absolute Potts model exhibits clear transitions into and out of ergodic dynamics 
\cite{hu2011phase} that coincide with the spin-glass type transitions found in the
(discrete Potts type) absolute Potts model.

\subsection{Crossovers in ``well-definedness'' of the planted state}

\begin{figure}
  \centering
  \begin{tabular}{cc}
    (a) &
    (b) \\
  \includegraphics[width=\halfwidth]{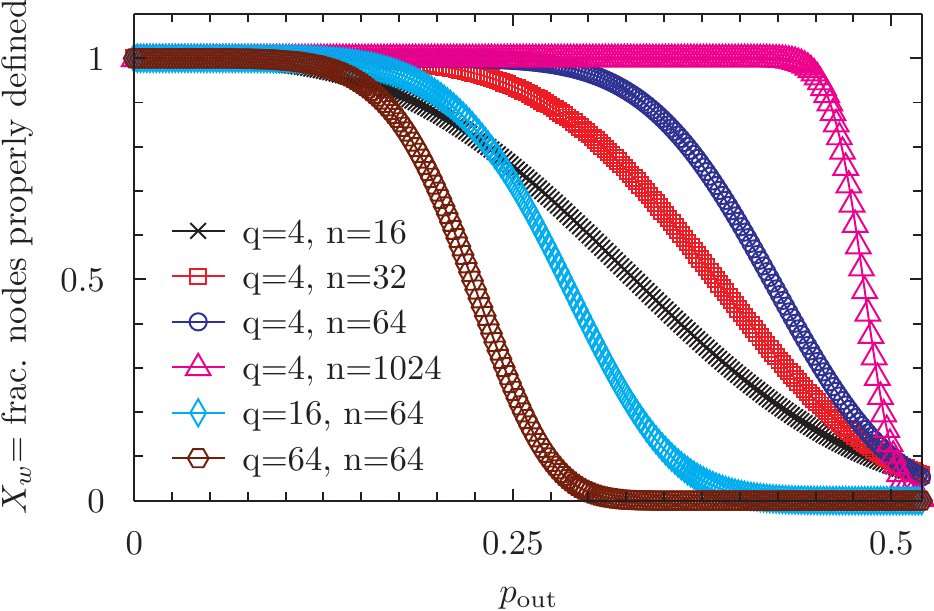} &
  \includegraphics[width=\halfwidth]{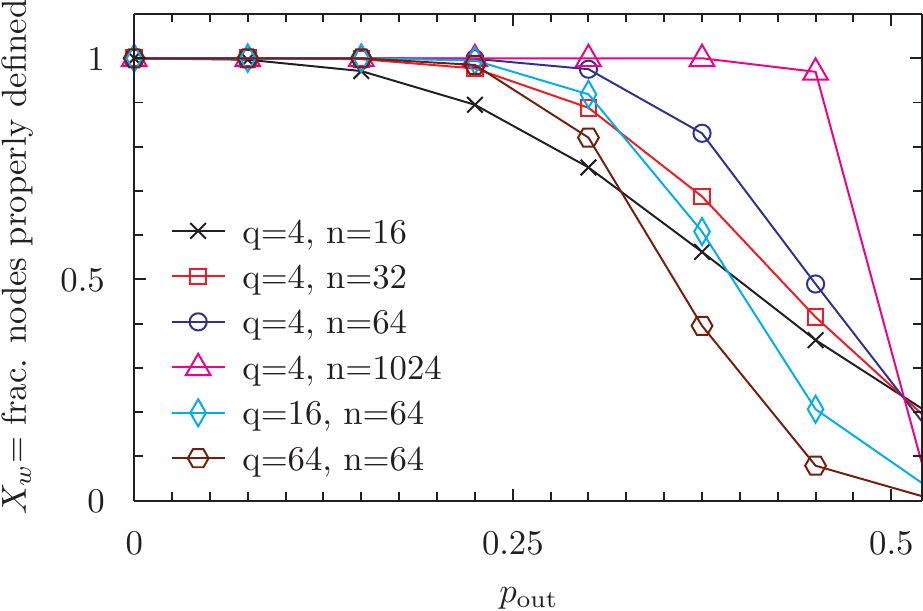}
  \end{tabular}
  \caption[Phase transitions in well-definedness]{The effects of varying $q$ and $n$ have at
    constant $p_\mathrm{in}=.5$.  We see that as $n$ decreases, we get a
    lower and lower $p_\mathrm{out}$ threshold for $X_w<1$, the bound
    for resolving communities.  As $q$ increases, our bound
    decreases.  Also, for any given $P_\mathrm{out}$, $X_w$ is lower
    for lower $n$ or greater $q$. (a) shows results for our
    independence approximation, (b) shows results for the correct
    numerical evaluation.  This illustrates how, as the community size $n$
    increases, a sharp phase transition between
    ill-defined and well-defined nodes is approached.}
  \label{fig:bmfitness-phasetransition}
\end{figure}

The current work investigates a related complementary problem- 
the viability of contending underlying ground
state (the best community assignments according to a particular CD
method) irrespective of how hard it may be to find such a ground state. 
In particular, we computed bounds of the fraction of well-defined nodes.
In a trivial solvable case, $X_{w}$ tends to unity while when the system is maximally disordered this bound veers from above towards $1/q$. These bounds {\it must not be naively confused and equated} with transitions appearing in various algorithmic approaches as studied in numerous earlier works. The ``well-definedness''
that we study in this work conveys information about the defined network structure.  When $X_w = 1$, all
nodes are properly defined in their communities and may be accurately
placed by an ideal CD algorithm.  For large graphs, as
$p_\mathrm{out}$ is increased, a cross-over occurs into a region where
nodes are not well defined according to their intended communities and
\emph{no} method should be able to properly detect \emph{all} of them
beyond the fraction given by $X_w$. We now explicitly discuss the behavior as the number of nodes grows large, see
Fig.~\ref{fig:bmfitness-phasetransition} in large dilute graphs ($n \gg 1$ and $k/n \ll 1$).  First, we examine the
independence approximation of Eq.~(\ref{eq:indepPfirst}).  When the
number of nodes per community $n\gg 1$, we may approximate $n \approx
n-1$ to obtain
\begin{equation}
\label{sharp}
  P^{\mathrm{ind}}_{w,AB} = \frac{1}{2} \left[
    1 - \mathrm{erf}\left(
      \frac{p_\mathrm{out} - p_\mathrm{in}}
           {\frac{1}{\sqrt{n}}\sqrt{2}\sqrt{p_\mathrm{in}(1-p_\mathrm{in})+p_\mathrm{out}(1-p_\mathrm{out})}}
      \right)
  \right].
\end{equation}
The denominator in this equation provides the width of the
erf decay.  This denominator decreases as $1/\sqrt{n}$ with increasing
community size $n$.  Thus, the greater the community sizes, the
sharper the crossover between the graph being well defined ($X_w
\approx 1$) and not ($X_w \ll 1$).  In the thermodynamic limit of
large $n$, a sharp change appears when the
parameter $(p_\mathrm{in}-p_\mathrm{out})=0$.
Fig.~\ref{fig:bmfitness-phasetransition} shows these effects.  Using
the exact expression of Eq.~(\ref{eq:Pexact}), we see how a finite $n$
crossover becomes progressively sharper as $n$ is increased.
Similar arguments will
apply to all structural or detectability transitions in graphs.
The manifest sharpness that emerges in the large $n$ limit of Eq. (\ref{sharp}) 
as ascertained in the well-definedness of viable ground states
goes hand in hand with the different yet complementary bona fide algorithmic thermodynamic phase transition 
discussed above and in earlier works. In \cite{hu2011phase}, as $n$ increased a spin-glass type transition emerged in a Potts type algorithm. We remark that in systems with finite $n$ yet divergent $N$, there is only a single undetectable phase \cite{hu2012stability,ronhovde2012global}. 

\section{Comparison with limits of real community detection algorithms}
\label{sec:NN_compare}

We now compare our result to the theoretical limit of detectability
(DKMZ limit)
in the infinite-size limiting
case\cite{decelle2011inference,nadakuditi2012graph}.
The DKMZ and NN result is in terms of the
variables $c_\mathrm{in}=Np_\mathrm{in}$ and $c_\mathrm{out}=Np_\mathrm{out}$, 
and $N= qn$.
Translating Eq.~(15) of NN \cite{nadakuditi2012graph} for $p_\mathrm{in}$
and $p_\mathrm{out}$, we have
\begin{equation}
  p_\mathrm{in}-p_\mathrm{out} = \sqrt{\frac{1}{n} \left(p_\mathrm{in} + (q-1)p_\mathrm{out}\right)  }.
  \label{eq:NNtransition}
\end{equation}
It is evident that as $n/q$ becomes large, the right side of this
equation approaches zero, forcing $p_\mathrm{in}=p_\mathrm{out}$ and
the results approaches the same limiting case of
$p_\mathrm{in}>p_\mathrm{out}$ for detectability of communities we had
in Eq.~(\ref{eq:palphabeta}).

Further, NN provide a formula for the fraction of nodes which can be
detected in SBMs via spectral-based methods\cite{nadakuditi2012graph}. For the $q=2$ case,
a pertinent parameter is set by
\begin{equation}
  \alpha^2 =
  \frac{ (c_\mathrm{in}-c_\mathrm{out})^2 - 2 (c_\mathrm{in}+c_\mathrm{out}) }
       { (c_\mathrm{in} - c_\mathrm{out})^2 }.
\end{equation}
The fraction $X_{DM}$ (fraction of nodes \emph{d}etected by
\emph{m}odularity) of correctly detected vertices using spectral methods that employ the modularity matrix is, according to NN, given by
\begin{equation}
  \label{eq:newman-detectable}
  X_{DM} = \frac{1}{2} \left[ 1 +
                \mathrm{erf}\left( \sqrt{\alpha^2 / 2 (1 - \alpha^2)} \right)
                      \right].
\end{equation}
We may use the same ideas as in the threshold
section (Sec. \ref{sec:thresholds}) to define
analogous thresholds
\begin{eqnarray}
  X_{DM}^{\mathrm{Th},h}\left( p_\mathrm{out}^{DM,h} \right) &=&
    1 - \frac{1}{N}, \\
  X_{DM}^{\mathrm{Th},l}\left( p_\mathrm{out}^{DM,l} \right) &=&
    \frac{1}{q} + \frac{1}{N}.
\end{eqnarray}

\begin{figure}
  \centering
  \includegraphics[width=\wholewidth]{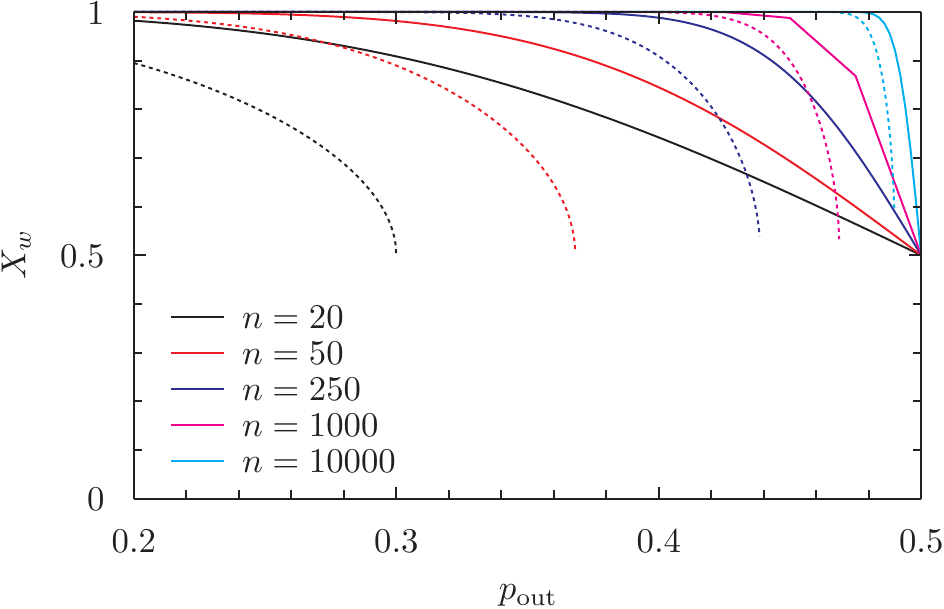}
  \caption[Well definedness vs other theoretical limits of
  detectability]{Stochastic block model $q=2$, $p_\mathrm{in}=.5$, and
    varying $n$.  Solid lines show $X_w$, dotted lines show the
    modularity-based fraction of nodes identified, with the right
    endpoints of the lines being the DKMZ
    limit of detectability.  As seen, in the
large $n$ limit, the fraction of particles well defined
    $X_w$ and the fraction classifiable via modularity converge.
The approach to this limit for small $n$ is different.}
  \label{fig:bmfitness-approach}
\end{figure}

In Fig.~\ref{fig:bmfitness-approach}, we observe how the NN fraction of
nodes detectable compares with the number of well-defined nodes by our
calculations.  We see that even in the ``accurate'' range of our
calculations ($X_w\approx1$), there is a significant difference
between the amount of present structure ($X_w$) and that detectable by
modularity ($X_{DM}$).  This divergence represents a region where
there is nominally structure present, yet this structure
cannot be detected.

\section{``Accurate'' methods agree with the $X_w$ curve}
\label{sec:BM-accurate_methods}

\begin{figure}
  \centering
  (a) \\
  \includegraphics[width=\twohighwidth]{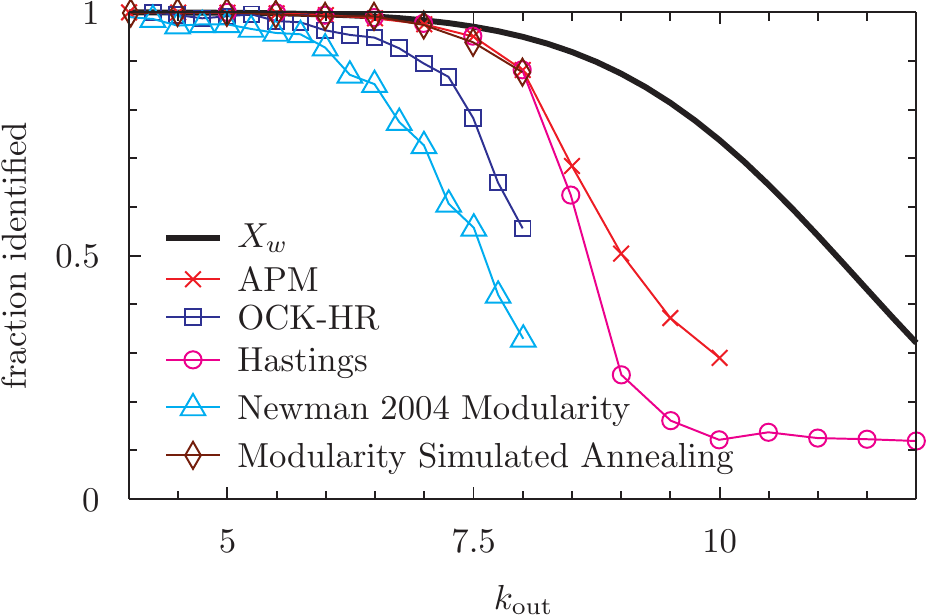}
  \\
  (b) detail of upper region \\
  \includegraphics[width=\twohighwidth]{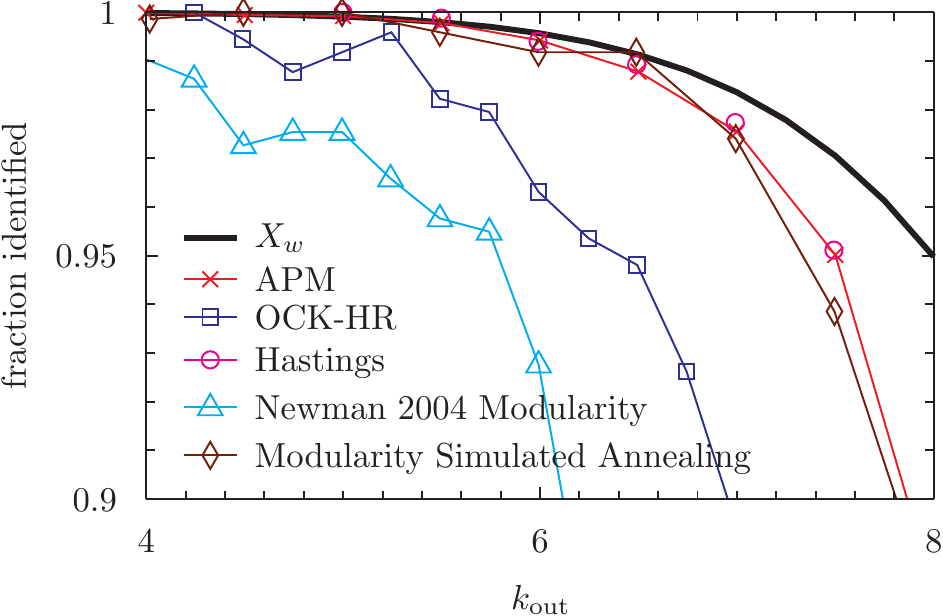}
  \\
  \caption[Demonstration of well-defined node fraction vs accurate
  community detection methods]{Demonstration of $X_w$ representing a limit for
    ``accurate'' community detection methods on the $q=4$, $n=32$ SBM
    graph with $\kin+\kout=16$.  We compare a variety of
    high-accuracy community detection methods (see text), and see that all
    methods are bound to detect less than $X_w$ nodes.  Panel (b) is a
    detail of (a), showing the high $X_w$ range.  When $X_w$ is high,
    we have minimal cascade effects, and $X_w$ is an accurate
    prediction of the accuracy of the most accurate methods.}
  \label{fig:bmfitness-methods-compare}
\end{figure}

In earlier sections, we derived a general bound for the fraction of nodes which are
properly connected to their communities. We expect \emph{any}
method to be constrained by these bounds.  In Fig.~\ref{fig:bmfitness-methods-compare},
we compare various accurate methods to our computed $X_w$.  As is seen,  the fraction of nodes that any method can detect is bounded from above by $X_w$.  Furthermore, the point
at which disparate methods begin losing accuracy is uniformly close to the same
value of $p_{out}$ at which $X_w$ is only slightly below one.  This indicates that, until the first nodes begin
being no longer well defined in their communities, it is easy for most
methods to accurately detect communities.

We compare the absolute Potts model (``APM'' above) \cite{ronhovde2010local} 
to a method based on desyncronized phase oscillators by Bocaletti {\it
  et. al.} (``OCK-HR'') \cite{boccaletti2007detecting},
Newman's original modularity optimization algorithm (``Newman 2004 Modularity'')
\cite{newman2004finding},
a modularity maximizing simulated annealing approach
by Danon {\it et. al.} (``Simulated Annealing'')
\cite{danon2005comparing},
and a belief propagation and mean field approach by
Hastings (``Hastings'') \cite{hastings74community}.  These are some of the
more common, and more accurate, community detection methods in
existence. We compare the experimental results of these methods with our computed $X_w$.
When $X_w$ is high, we observe that the most accurate algorithms are
able to \emph{almost exactly} detect a fraction $X_w$ fraction of the nodes.
This indicates that not only is $X_w$ a bound, but is a fairly
complete calculation for the region without cascade effects.  When we
move to higher $p_\mathrm{out}$, our ability to detect communities
diverges from the $X_w$ theoretical limit.  This can be due to cascade
effects leading to inaccurate calculation of $X_w$, as described in
Sec.~\ref{fraction}, where each ill-defined node affects more nodes
than just itself.  Alternatively, the divergence of $X_w$ and
fractions of nodes detectable, as described in
Sec. \ref{sec:NN_compare}, could also be a cause
for this divergence.

\section{General considerations applicable to problems other than the stochastic block model}
\label{general_section}

In what follows, we briefly discuss rather trivial extensions that
enable us to compare and extend some considerations to
computational/satisfiability problems other than community
detection as applied to the stochastic block model.  We first express
our conditions for well defined community detections as a requirement
that a certain cost function vanish. We then further discuss a trivial
yet general relation between computational problems with
planted solutions and the viability of finding an optimal solution for
the computational or satisfiability problems. We finally remark on
non-rigorous bounds on the extent of the hard phase in the SBM  
and possible extensions to other computational problems.

\subsection{Effective energy}

In many computational problems, a certain cost function is to be
minimized. All of our analysis thus far has focused on when
Eq.~(\ref{important_density}) may be satisfied.  This led to our
expressions for the fraction of correctly identified nodes $X_w$. We
may restate the condition for well defined communities of
Eq.~(\ref{important_density}) by constructing an energy function
\begin{eqnarray}
\label{Hamiltonian}
E = \sum_{A=1}^{q} \sum_{a \in A} \Theta \Big[ \max_{B \neq A} \left\{ \rho_{\mathrm{out},B}^{a}\right\} - \rho_\mathrm{in}^{a} \Big].
\end{eqnarray}
The above sum is over all communities $A$ and all nodes $a$ within them and $\Theta(x)$
the Heaviside function ($\Theta(x>0) = 1$, $\Theta(x<0) =0$). We may now
ask whether there exist a partition (or partitions) for
which the energy $E=0$.  As the energy function of Eq. (\ref{Hamiltonian}) counts the number of
nodes not satisfying our well-defined criteria, we have the correspondence
\begin{equation}
  E = N(1 - X_w).
\end{equation}
According to the postulates regarding well-definedness in this work,
\emph{minimizing this energy corresponds to a rudimentary form of
  community detection}.  
With the Hamiltonian at hand, we can go beyond an analysis of
the system ground states and examine whether it is possible to optimally
satisfy the community detection criteria. We can now, as in \cite{hu2011phase}, broadly define and examine finite temperature entropy, energy, spin-glass type phase transitions that they exhibit, and much more.

\subsection{Planted states as variational states in general graphs}

A trivial yet important point which we wish to emphasize is the
following. The planted graph partition might be viewed as a
variational state. That is, if the planted partition satisfies
Eq.~(\ref{important_density}) or, correspondingly, is a zero energy
(ground state) of the energy function of Eq.~(\ref{Hamiltonian}) then
clearly there is at least one state for which
Eq.~(\ref{important_density}) is satisfied.  Similarly, if a planted
state violates a certain number of conditions of the form of
Eq.~(\ref{important_density}) then there exists at least one partition
(i.e., the planted state) which violates the same number of
conditions.  However, by adjusting the community assignments, we may
find a different state (i.e., one which differs from the planted state)
which break a smaller number of these conditions and is thus of lower energy.  We will denote the fraction
of well defined nodes and associated threshold values for this minimal energy state(s) by $\overline{X}_w$ and
$\overline{p}^{l}$.  This is primarily important when
we are detecting communities without knowledge of the planted state (i.e., the general practical task of community detection
algorithms). In such a case, we search for a state that violates the
least number of constraints of the form of
Eq.~(\ref{important_density}) or equivalently has the lowest energy in
Eq.~(\ref{Hamiltonian}). 

Thus, for any given graph, we will trivially have the inequalities 
\begin{eqnarray}
  \overline{p}_\mathrm{out}^{h} &\ge& p_\mathrm{out}^{h}, \\
  \overline{p}_\mathrm{out}^{l} &\ge& p_\mathrm{out}^{l}.
\label{trivial_threshold1}
\end{eqnarray}
Similar to Eqs.~(\ref{upper_threshold}, \ref{lower_threshold}),
on the lefthand side of Eq. (\ref{trivial_threshold1}),  $ \overline{p}_\mathrm{out}^{h},   \overline{p}_\mathrm{out}^{l}$
denote, respectively, values of the noise at which $X_{w}$ is equal to the lower and upper
threshold values for the correctly placed nodes relative to the lowest energy state (with no knowledge of a planted state).  
By contrast, on the righthand side of Eq. (\ref{trivial_threshold1}),
$p_\mathrm{out}^{h}, p_\mathrm{out}^{h}$ denote the noise values $p_{out}$ at which the fraction 
of correctly placed nodes $X_{w}$ achieves lower and upper
threshold values relative to a known planted state. 

The variational state provided by the planted state, if anything, may lead us to
believe that our threshold is \emph{smaller} than it actually is for finding sensible
community partitions.
The fact that we detect the lowest-energy ground state, instead of the
variational (planted) state, leads us to infer a \emph{greater}
threshold for community detection than we actually have.  This may
be of utility when the true
communities are not known, and the accuracy of the detection must be
inferred from the relative noise.

\subsection{Possible proof of principle on boundaries of hard phase in which problems cannot be easily solvable yet
for which contending solutions can be polynomially checked}

As is evident from Fig. \ref{fig:bmfitness-2x50}, there is (for
general non-vanishing $q/n$ and/or in dense graphs) a finite interval
of $p_{out}$ values for which (i) {\it there are} on average, as we
proved, still meaningful partitions (which can be checked in
polynomial time for the conditions specified by
Eq. (\ref{important_density})) yet (ii) according to, e.g., 
%\cite{decelle2011inference,nadakuditi2012graph,mossel2012stochastic}
\cite{nadakuditi2012graph}
{\it there are no spectral algorithms} that can efficiently ascertain
structure. Together, this suggests as a matter of principle a route
for establishing a hard phase as probed by various specific algorithms. The hard phase consists of very
challenging graph partitioning problems that cannot be solved
efficiently (i.e., may be non-polynomial problems) by any algorithm yet purported
solutions can be checked in polynomial time (i.e., belonging to
NP). Namely, this situation is one for which the old conjecture $ P                     
\neq NP$ may explicitly come to life\cite{fortnow2009status}. We
caution that our work (which leads to item(i) above) centered on the
use of Eq. (\ref{important_density}) while others
\cite{decelle2011inference,nadakuditi2012graph,mossel2012stochastic}
did not use these criteria for their point of departure. 
When satisfied, the criteria of Eq. (\ref{important_density}) suggest
community structure yet it is possible that non-trivial meaningful clusters
can be inferred such that they adhere to other criteria. The rigorous
upper bounds that
we derived in the current work do not incorporate ``cascade effects''
(Sec.~\ref{fraction}); when these are taken into account, cascade effects may generally 
lead to lower threshold values of the noise beyond which well defined structure
ceases to exist. We reiterate that the cavity-type approximations of
DKMZ
\cite{decelle2011inference} lead to results identical to those of NN
as suggested by spectral methods. We should further remark anew that,
as illustrated by \cite{mossel2012stochastic}, no Bayesian inference
algorithm can detect structure beyond for noise values larger the DKMZ expression for the
special case of sparse SBM graphs with $q=2$ communities.  
The general considerations that we invoked here for finding where the hard
phase may potentially appear may be replicated to other computational problems.

\section{Conclusions}
\label{sec:BM-conclusions}

We conclude with a brief synopsis of our results:

$\bullet$ \emph{Community detection as a function of graph structure.}  Detecting communities in general graphs is an 
NP-type problem that has
gained much attention in the last decades. More recently, several groups have examined a particular subclass - the stochastic block model graphs - with the goal of
calculating noise thresholds on the
ability to detect community structure via various algorithms such as those involving spectral methods or 
considerations related to the fundamental ability of disparate methods to infer structure. These methods were 
examined elegantly via the cavity-type approximations and have been bolstered by other considerations. In this work, 
we take a different path to attack this problem. 
Specifically, instead of studying limitations of various algorithmic or inference approaches we have turned the problem
around and examined the
properties of the graph itself to examine to what the community detection partition of the system may have a well-defined solution. {\it This approach enabled us to derive universal bounds
independent of any particular community detection algorithm and/or inference methods/approximations}.  We invoked
a simple criterion for community structure that relies on edge densities. 
Using this, we derived a relationship for the fraction of nodes consistent with the correct
community assignment.  Our approach is much simpler that of past
works and offers a complimentary understanding on the limits of detectability. The ideas introduced 
in our work, with the principle of focusing on the problem itself, independent of any known algorithms or inference approximations, might have also applications in the analysis of 
other hard computational problems.

$\bullet$ \emph{Rigorous bounds on well-definedness and community detection
  algorithms.}  Our bound on the highest number of correctly identifiable vertices of
  the planted state, in the non-sparse case (such that the exact binomial distribution 
  may be replaced by a normal distibution) is given by $X_w(p_\mathrm{in}, p_\mathrm{out}, q, n)$ 
  of Eq. (\ref{xweqe}) 
  (wherein the corresponding CDF  is given by Eqs. (\ref{eq:indepPnormalapprox}, \ref{pwab})
and the PDF is given by Eq. (\ref{inp})). We reiterate that this
provides a strict upper bound on the accuracy of any community detection
method for planted equal community size stochastic block model graphs.
For sparse graphs, the exact binomial distribution (or its Poisson distribution approximation)
may be invoked in the probability distribution functions. 
Furthermore, we have derived an ``independence approximation'' which is
most accurate for a small number of communities $q$ or high $X_w$.  We
have shown, by a simple application of Jensen's inequality, that the independence approximation leads to a strict lower bound
on the actual fraction of well defined nodes, $X_w^{\mathrm{ind}} \le X_w$.
Finally, by comparison with previous accurate community
detection algorithms, we have shown that our $X_w$ bound is indeed an
upper limit for all of these algorithms as is clearly seen in 
Fig.(\ref{fig:bmfitness-methods-compare}). 
We achieve the greatest
accuracy when $X_w$ is high, minimizing cascade effects of
ill-definedness. We have established that by focusing on ill-defined nodes
and assigning these to the optimal communities, the transition to 
structureless partitions is no longer as precipitous as it is otherwise.
The bounds that we obtain on $X_w$ correlate emulates and narrowly lie above the curves found by NN \cite{nadakuditi2012graph}
for the fraction of well-defined nodes.

$\bullet$ \emph{Sharp behavior in large-size limit.}  We have demonstrated
that as community size $n$ increases, the width of the transition
between well-defined and ill-defined decreases.
The absolute Potts model approach to community detection
problems \cite{ronhovde2010local,ronhovde2009multiresolution}
substantiated how the width of
intermediate ``hard'' phase
decreases as $n/q$ becomes progressively larger\cite{hu2011phase,
  hu2011replica, hu2012stability, ronhovde2012global} for dilute
graphs.

 $\bullet$ \emph{Difference between solvable problems and checkable solutions.}
 Taken together, the results that we derived in this work for
 (i) cases when graphs have well defined underlying structure 
 and (iii) earlier results \cite{nadakuditi2012graph,decelle2011inference}
 concerning the limitations of general spectral algorithms and inference approaches 
 suggest bounds on the hard phase.
 Within the hard phase, the community detection problem
 cannot be efficiently solved yet for which purported solutions can
 be checked in polynomial time. Away from the large $n/q$ limit in 
 dilute graphs the combined results of (i) and (ii) allow for the emergence of polynomially checkable
 yet extremely hard to solve problems. That is, away from these limits,
the hard phase may appear. The surplus of the fraction of well-defined nodes 
$X_{w}$ that we found in the current work (irrespective of applied algorithm)
as compared to the fraction found by the modularity matrix based algorithm
of NN \cite{nadakuditi2012graph} is notable. This disparity may highlight difference between
solvable problems (NN) and rigorous bounds on proposed contending checkable solutions (the current work) by examining the fraction of nodes that satisfy the criteria for well-definedess.

$\bullet$ \emph{Relevance to benchmark graphs.}  Perhaps the most practical
implication of this work relates to the construction and analysis of
benchmark graphs for community detection.  In order to judge the
effectiveness of any algorithm, one must know the expected maximal possible
performance.  We investigated the performance of various algorithms and 
examined upper and lower threshold values of noise
in stochastic block model graph benchmarks and the role
of ill-defined nodes where the community detection criteria are 
not satisfied.  Benchmark graphs such as that of LFR \cite{lancichinetti2008benchmark,lancichinetti2009benchmarks} 
can be designed to avoid ill-defined nodes by
applying a ``rewiring'' step which keeps internal to external edges at
as constant a ratio as possible. In a companion work, we advanced {\it a general edge density based approach to community
detection} which complements the edge density criteria of Eq. (\ref{important_density}) that we invoked
in the current work\cite{darst2013edge}. Aside from examining nodes
and their respective edge densities as criteria for stable communities
as we have in this work, we may
also apply similar criteria to the {\it density of connections between links} (i.e., look at
a dual graph formed by the vertices placed at the centers of each link 
of the original graph and ask whether 
an edge density criterion is satisfied for the edges between these
link centers) {\it or connections between triangles}, 
etc. \cite{darst2013edge}.  

Our analysis, while detailed, raises many further questions.  Can this
analysis be extended to different cost functions, or graphs with power
law degree distributions?  Can we successfully model cascade effects,
where each incorrect nodes affects the size of communities and thus
affects more than just itself?  Perhaps most importantly, how does
this issue intersect with real-world graphs?  We hope that in such
graphs with real-world planted states, there is some best community
definition.  Using an analysis similar to the one presented here, to
what degree does the graphs edge structure reflects those communities?

In closing, we remark that {\it the analysis we employ here could, potentially, be extended and applied to other
random computational problems}. In addition to tackling the algorithmic limits
of the solution process, our approach enables one to examine said
limits from structural standpoint of the problem itself.

\textbf{Note added in proof}: The bulk of this work first appeared in
the PhD thesis of one of us (RKD) in October 2012 (available online at \cite{darst2012lattice}),
which extensively studied the issues surrounding ill-defined
nodes.  We very recently became aware of a
related preprint \cite{floretta2013stochastic} which shares some
features and a viewpoint similar to our work.

\begin{acknowledgements}
  We would like to thank Santo Fortunato, Cris Moore, and Dandan Hu
  for useful discussions.  RKD would like to thank the John and Fannie
  Hertz Foundation for research support via a Hertz Foundation
  Graduate Fellowship.  The work at Washington University in St Louis
  has been supported by the National Science Foundation under NSF
  Grant DMR-1106293.  ZN also thanks the Aspen Center for Physics for hospitality and
  NSF Grant \#1066293.
\end{acknowledgements}

\appendix
\section{Probability distribution nomenclature}
\label{sec:probdistribution}

Below, for the sake of clarity, we briefly make explicit the very
standard shorthand notations of PDF and CDF that we employ.
$\mathrm{PDF}\left[\mathbb{X}\right]$ represents the
\emph{probability density function}, the probability of the
random variable $\mathbb{X}$ being in the infinitesimal interval $[x,
x+dx]$.  In Eq. (\ref{eq:Pexact}), this PDF associated with $ \rho_{\mathrm{in},A}$ is given by
Eq. (\ref{inp}).  Generally, for any distribution, the PDF is given by the respective integral
\begin{equation}
  \label{eq:PDF}
  \int_a^b \mathrm{PDF}\left[ \mathbb{X} \right](x) dx = P\left[ a
    \leq \mathbb{X} \leq b \right].
\end{equation}
The \emph{cumulative distribution function}
($\mathrm{CDF}$), for a general distribution function, corresponds to the probability of the random variate
$\mathbb{X}$ being less than or equal to a given value,
\begin{equation}
  \mathrm{CDF}\left[ \mathbb{X} \right](b)
  = \int_{-\infty}^b \mathrm{PDF}\left[ \mathbb{X} \right](x) dx
  = P\left[ \mathbb{X} \leq b \right].
  \label{eq:CDF}
\end{equation}
For the external links with a PDF given by Eq. (\ref{outp}), the corresponding CDF is [as we stated
in the main text] given by Eqs. (\ref{eq:indepPnormalapprox}, \ref{pwab}). 
Both the PDF and CDF may be associated with either the original discrete problem (described by a binomial distribution) or
its continuous approximation (a Gaussian as in Eq. (\protect{\ref{eq:pdifftransition}})), with the corresponding trivial change between
integrals to sums in the equations above if the discrete form is sought.

\end{document}